\documentclass[pra,showpacs,showkeys,english]{revtex4}
\usepackage{indentfirst}
\usepackage[T1]{fontenc}
\usepackage[latin9]{inputenc}
\usepackage{color}
\usepackage{babel}
\usepackage{amsthm}
\usepackage{amsmath}
\usepackage{graphicx}
\usepackage{amssymb}

\makeatletter

\theoremstyle{plain}

\newcounter{thm}
\newcounter{lem}
\newcounter{rmk}

\begin{document}

%\begin{CJK*}{GBK}{kai}

\title{Optimum Unambiguous Discrimination of Linearly Independent Pure States}

\author{Shengshi Pang$^{1}$}

\author{Shengjun Wu$^{1,2}$}

\affiliation{$^{1}$Hefei National Laboratory for Physical Sciences at
Microscale, University of Science and Technology of China, Hefei,
Anhui 230026, China \\
$^{2}$Lundbeck Foundation Theoretical Center for Quantum System
Research, Department of Physics and Astronomy, Aarhus University,
DK-8000 Aarhus C, Denmark}

\date{\today}

\begin{abstract}
Given $n$ linearly independent pure states and their prior probabilities,
we study the optimum unambiguous state discrimination problem. We derive the conditions for
the optimum measurement strategy to achieve the maximum average success probability,
and establish two sets of new equations that must be satisfied by the optimum solution in different situations.
We also provide the detailed steps to find the optimum measurement strategy.
The method and results we obtain are given a geometrical illustration with a numerical example.
Furthermore, using these new equations, we derive a formula which shows a clear analytical relation between the optimum solution and the $n$ states to be discriminated. We also solve a generalized equal-probability measurement problem analytically.
Finally, as another application of our result, the unambiguous discrimination problem of three pure states is studied in detail and analytical solutions are obtained for some interesting cases.
\end{abstract}
\pacs{03.67.-a, 03.65.Wj, 03.65.Ta, 42.50.Dv}% PACS, the Physics and Astronomy % Classification Scheme.
\keywords{Unambiguous state discrimination, optimum measurement strategy}
\maketitle
%\end{CJK*}

\section{Introduction\label{sec:Introduction}}

Discrimination of quantum states has been an interesting and attractive
problem in quantum information science for a relatively long time
\cite{Helstrom,Holevo 1,Holevo 2}. Since deterministic and error-free
discrimination of an arbitrary set of quantum states is generally
impossible due to the basic principles of quantum mechanics, just
like other "no-go" theorems \cite{no-cloning nature 1982,no deleting 2000,quantum cloning schodinger's cat 2000},
the problem of finding an effective scheme of state discrimination  has attracted a lot of attention and has played an import role
in the study of quantum communication and cryptography. Considerable work has been
devoted to this problem and it has developed rapidly recently.

The task of state discrimination is to discriminate the state of a quantum system from a given finite set of possible states with certain prior probabilities, and there are mainly two kinds of strategies to complete this task. One kind of strategy is called \emph{minimum
error discrimination} \cite{Holevo minimum error,Helstrom,ieee minimum error}
which requires that the average probability of identifying a wrong
state is minimized. There have been numerous results for this kind
of discrimination strategy\cite{mini error 3 mirror symmetric ,mini error c l chou,mini error ieee 2004,minimum error symmetric,mirror symmectric mini error,mixed mini error 2008,optimal minimum error,mini error subset},
some of which are quite interesting, like the weighted
square-root measurement \cite{sqrt root minimum error,sqrt root 2001,sqrt root 1,sqrt root 2,sqrt root 3,sqrt root 4,sqrt root 5,sqrt root 6}
and that measurement sometimes does not aid in discriminating
certain set of states \cite{measurement not aid discrimination,a family of measurement not aid}.

Another important kind of discrimination strategy is \emph{unambiguous discrimination}
pioneered by \cite{Ivanovic,Dieks,Peres}, which requires that no
error occurs in the identification of the states at the expense
of obtaining an inclusive result with some non-zero probability. A
lot of research has also been performed on this kind of discrimination
strategy \cite{jaeger unambiguous,chefles unambiguous 1,mathematical nature of ,peres unambiguous 1998,set discrimination,set discrimination 2,duan lu ming,08 nian,determinant 0},
including unambiguous discrimination of symmetric states
\cite{symmetric unambiguous } and unambiguous discrimination between
mixed states \cite{mixed unambiguous 1,mixed and pure unambiguous 2,mixed unambiguous 3,mixed unambiguous 4,mixed unambiguous 5,mixed unambiguous 6,mixed unambiguous 7,mixed unambiguous 8,mixed unambiguous 9,mixed unambiguous 10}.

In addition to the above two strategies, research has also been performed
on mixed strategies involving minimum-error discrimination and unambiguous
discrimination together \cite{mixed stragety 1,mixed strategy 2,mixed strategy 3,mixed strategy 4},
in order to achieve balance between the accuracy and the efficiency
of state identification. It is, in general, very difficult to maximize
the average probability of successfully discriminating the given states analytically,
but some special techniques such as semidefinite programming have been
employed to solve this computation problem numerically \cite{semidefi 0,semidefi 1,ieee epm,08 nian}.

In this article we shall study the properties of the optimum strategy
for unambiguous discrimination of $n$ pure states, and give a detailed method to obtain
such an optimum strategy. According to \cite{chefles unambiguous 1,duan lu ming},
one can manage to discriminate a set of states unambiguously with non-zero
success probabilities if and only if the given states are linearly
independent. We shall hold this assumption throughout this article.
The main method we use for studying this problem is to put the individual probabilities that each state is successfully identified together as a vector in the $n$-dimensional real space $\mathbb{R}^{n}$ (and do the same to the $n$ prior probabilities), and study the properties of the minimum eigenvalue of the matrix $X-\Gamma$  under the optimum strategy by vector analysis techniques ($X$ is the Gram matrix of the $n$ states to be discriminated and $\Gamma$ is the diagonal matrix with the success probabilities as its diagonal elements, they will be defined explicitly later in Theorem \ref{positivity}). From the properties of the minimum eigenvalue of $X-
\Gamma$, we shall establish two sets of equations that the optimum solution of the unambiguous discrimination problem must satisfy in different situations, and these equations will turn out to have some intuitive geometrical meanings.

This article is organized as follows. In Sec. \ref{sec:Problem-Description:-POVM}
we shall give a general description of the unambiguous state discrimination
problem that we are interested in and the POVM formalism that we shall
use in this article. In Sec. \ref{sec:Exploration of optimal point},
we derive the properties of the optimum measurement strategy and the new
equations that can be used to work out the optimum solution. Examples are
also given to geometrically and numerically illustrate
our method of solving the unambiguous discrimination problem. Sec.
\ref{sec:Maximum-Average-Success} is devoted to deriving an analytical
formula which characterizes a simple relation between the maximum
average success probability and the $n$ pure states to be unambiguously
discriminated. In Sec. \ref{sec:Generalization-of-Equal-Probability},
a generalized version of equal-probability-measurement (EPM) problem
\cite{ieee epm} is studied and an analytical solution is obtained.
Finally we apply the results obtained in Sec. \ref{sec:Exploration of optimal point}
to the case of three linearly independent pure states in Sec. \ref{sec:Special-Case:}
and work out analytical solutions for some interesting cases.

\section{Problem description and POVM formalism\label{sec:Problem-Description:-POVM}}

In this article, our problem is how to unambiguously identify the state
of a quantum system with the maximum average success probability since the discrimination of non-orthogonal states is generally probabilistic.
And all we know is that this state belongs to
a given set of $n$ linearly independent states $\left\{ |\psi_{i}\rangle\right\} _{i=1}^{n}$
with given prior probabilities $\gamma_{i}$ ($i=1,\cdots,n$).
We would like to obtain some analytical conditions that the optimum solution should satisfy, and provide a detailed method for obtaining the optimum solution of the problem.

In the following study of the problem, the $n$ prior probabilities will be
denoted by a real vector $\boldsymbol{\gamma}$ in the space $\mathbb{R}^{n}$
and the success probabilities $p_{i}$ ($i=1,\cdots,n$) for unambiguous outcomes
will also be denoted as a real vector $\boldsymbol{p}$ in $\mathbb{R}^{n}$
for short. The average success probability for unambiguous discrimination
can be written as $\bar{p}=\sum_{i=1}^{n}\gamma_{i}p_{i}\boldsymbol{=\gamma}\cdot\boldsymbol{p}$ then.

Since only the probabilities of the measurement outputs are concerned in this problem, we shall use the POVM (\emph{positive-operator-valued
measure}) formalism \cite{povm} which is a good description for the statistics of a general physics process. A POVM consists of a set of \emph{POVM elements $\left\{ \Pi_{i}\right\} $}
satisfying $\Pi_{i}\geq0$ and $\sum_{i}\Pi_{i}=I$, where $I$ represents
the identity operator (or matrix). A Hermitian operator is said
to be \emph{positive} and denoted by "$\geq0$" if all of its eigenvalues
are non-negative. To unambiguously identify the state, we require
that \begin{equation}
p(j|i)=\langle\psi_{i}|\Pi_{j}|\psi_{i}\rangle=p_{i}\delta_{ij},\;\forall i,j=1,\cdots,n\label{eq:-25}\end{equation}
where $p(j|i)$ represents the probability of obtaining
the result $j$ when the original state of the system is actually
$|\psi_{i}\rangle$ and $p_{i}$ denotes the probability of
correctly identifying $|\psi_{i}\rangle$.

Let $\mathcal{H}$ denote the $n$-dimensional Hilbert space spanned by the given
set of linearly independent pure states $\left\{ |\psi_{i}\rangle\right\} _{i=1}^{n}$.
The total Hilbert space of the system, denoted by $\mathcal{H}_{total}=\mathcal{H}\oplus\mathcal{H}^{\perp}$,
may be larger than $\mathcal{H}$, where $\mathcal{H}^{\perp}$ denotes
the subspace orthogonal to $\mathcal{H}$. Any operator acting on
$\mathcal{H}_{total}$ is equivalent to an operator on $\mathcal{H}$
by projecting onto the space $\mathcal{H}$ when only
the effect on $\mathcal{H}$ is considered. Therefore, without loss of generality,
we shall restrict our POVM elements to those acting on the space $\mathcal{H}$
spanned by the given set of states. In order to discriminate the states
$\left\{ |\psi_{i}\rangle\right\} _{i=1}^{n}$ unambiguously, the
POVM element $\Pi_{i}$ that identifies the $i$th state $|\psi_{i}\rangle$
must be orthogonal to the subspace spanned by the other $n-1$ states
according to Eq. \eqref{eq:-25}, therefore the rank of $\Pi_{i}$
should be no larger than $1$, for all $i=1,\cdots,n$. Thus each
POVM element that successfully identifies a certain state should have
the form \cite{08 nian,ieee epm}
\begin{equation}
\Pi_{i}=p_{i}|\widetilde{\psi_{i}}\rangle\langle\widetilde{\psi_{i}}|,\label{eq:-11}
\end{equation}
where $p_i$ is the success probability to identify the $i$th state and $|\widetilde{\psi_{i}}\rangle$ is an unnormalized state orthogonal
to $|\psi_{j}\rangle$ for all $j\neq i$.
It can be seen by substituting Eq. \eqref{eq:-11} into \eqref{eq:-25} that $\bigl|\langle\psi_{i}|\widetilde{\psi_{i}}\rangle\bigr|^2=1$ and there is freedom for each  $|\widetilde{\psi_{i}}\rangle$ to have an arbitrary phase, and without loss of generality we make a specific choice of the phase such that
\begin{equation}
\langle\psi_{j}|\widetilde{\psi_{i}}\rangle=\delta_{ij},\label{eq:92}
\end{equation} for convenience.

Let $\Phi$ denote the matrix with $|\psi_{i}\rangle$ as its $i$th
column. Define $\widetilde{\Phi}$ as\begin{equation}
\widetilde{\Phi}=\Phi(\Phi^{\dagger}\Phi)^{-1},\label{eq:0}\end{equation}
then\begin{equation}
\Phi^{\dagger}\widetilde{\Phi}=I.\label{eq:91}\end{equation}
Note that Eq. \eqref{eq:0} cannot be simplified to $\widetilde{\Phi}=(\Phi^{\dagger})^{-1}$ in general, because the states $\left\{ |\psi_{i}\rangle\right\} _{i=1}^{n}$ may belong to a larger Hilbert space of which the dimension is greater than $n$, implying that the matrix $\Phi$ may not be a square matrix. (In the last paragraph, we only restrict the POVM elements to be those that act on the space spanned by these states, but we do not make any restrictions on the representation of these states.)
Comparing \eqref{eq:92} and \eqref{eq:91}, we know that $|\widetilde{\psi_{i}}\rangle$
is exactly given by the $i$th column of the matrix $\widetilde{\Phi}$.

Since unambiguous discrimination of the states $\left\{ |\psi_{i}\rangle\right\} _{i=1}^{n}$
is probabilistic if they are not orthogonal to each other, there exists
a POVM element $\Pi_{0}$ which gives the inconclusive result, and it
can be written as\begin{equation}
\Pi_{0}=I-\sum_{i=1}^{n}p_{i}|\widetilde{\psi_{i}}\rangle\langle\widetilde{\psi_{i}}|.\label{eq:-9}\end{equation}
Since any POVM element must be positive to represent a physically
realizable process, it is required that\begin{equation}
\Pi_{0}=I-\sum_{i=1}^{n}p_{i}|\widetilde{\psi_{i}}\rangle\langle\widetilde{\psi_{i}}|\geq0.\label{eq:-10}\end{equation}
This positivity inequality is an essential constraint on the unambiguous
discrimination scheme and is the starting point of the discussions
in this article.

When the set of states $\left\{ |\psi_{i}\rangle\right\} _{i=1}^{n}$
and prior probabilities $\gamma_{i}$ ($i=1,\cdots,n$) are given,
the POVM elements for the measurement ($\Pi_{i}=p_{i}|\widetilde{\psi_{i}}\rangle\langle\widetilde{\psi_{i}}|$,
$\Pi_{0}=I-\sum_{i=1}^{n}\Pi_{i}$) depend only on the variables
$p_{i}$, i.e. the success probabilities, since $|\widetilde{\psi_{i}}\rangle$
($i=1,\cdots,n$) can be determined and explicitly given by the
$i$th column of $\widetilde{\Phi}$  defined in \eqref{eq:0}.
Therefore, searching for the optimum solution is to find a set of
success probabilities $p_{i}$ (corresponding to a point or a vector
$\boldsymbol{p}$ in $\mathbb{R}^{n}$) such that their weighted average
(with the prior probabilities $\gamma_{i}$ as the weights) is maximized
under the restrictions $p_{i}\geq0$ ($i=1,\cdots,n$) and $\Pi_{0}\geq0$. The optimum
solution is denoted by $\boldsymbol{p}_{opt}$, and shall be called \emph{
optimum point} sometimes throughout the rest of the article; and the main goal
of this article is to find the optimum point $\boldsymbol{p}_{opt}$.

\section{Exploration of the optimum strategy\label{sec:Exploration of optimal point}}

In this section, we shall obtain some properties of the optimum strategy
for the unambiguous state discrimination problem, and provide a systematic
way to obtain the maximum average success probability and the optimum
measurement strategy.

\subsection{General properties and methods\label{sub:General-Situations}}

As the first step to study the optimum unambiguous state discrimination
problem described in Sec. \ref{sec:Problem-Description:-POVM}, we
are going to re-derive the positivity condition given by Duan and
Guo \cite{duan lu ming} in a more concise way, using the POVM representation,
and prove a convexity property of the set of all feasible $\boldsymbol{p}$'s.

\refstepcounter{thm}\label{positivity}\emph{Theorem \thethm (Positivity and Convexity).} Suppose $\left\{ |\psi_{i}\rangle\right\} _{i=1}^{n}$
is a set of linearly independent pure states. Let $X=\Phi^{\dagger}\Phi$
where $\Phi$ is the matrix whose $i$th column is $|\psi_{i}\rangle$
$\left(i=1,\cdots n\right)$, and $\Gamma=\text{diag}(p_1,\cdots,p_n)$
where \textbf{$p_{i}$} is the success probability to unambiguously
discriminate $|\psi_{i}\rangle$ ($i=1,\cdots,n$). Then i)
\begin{equation}
X-\Gamma\geq0,\quad
\Gamma\geq0.\label{eq:1}\end{equation}
ii) Let $\mathcal{S}$ denote the set of points $\boldsymbol{p}$ satisfying
the positivity condition \eqref{eq:1}, then $\mathcal{S}$ is convex.

\emph{Proof.} i) Using the definition of $\Gamma$ in this theorem,
Eq. \eqref{eq:-10} can be re-written as
\begin{equation}
I-\widetilde{\Phi}\Gamma\widetilde{\Phi}^{\dagger}\geq0.\label{eq:-17}
\end{equation}
Substituting Eq. \eqref{eq:0} into \eqref{eq:-17}, we have
\begin{equation}
I-\Phi(\Phi^{\dagger}\Phi)^{-1}\Gamma(\Phi^{\dagger}\Phi)^{-1}\Phi^{\dagger}\geq0.\label{eq:-18}
\end{equation}
According to the property of positive matrix, we can multiply Eq.
\eqref{eq:-18} by $\Phi^{\dagger}$ from the left side and by $\Phi$
from right side, therefore immediately get the first inequality in
Eq. \eqref{eq:1}.
The second inequality of Eq. \eqref{eq:1} must be satisfied since the success probabilities
$p_{i}\geq0$ ($i=1,\cdots,n$) must be non-negative.

ii) Let $\boldsymbol{p}_{1}$ and $\boldsymbol{p}_{2}$ denote two
arbitrary points in $\mathcal{S}$, and $\Gamma_{1}$, $\Gamma_{2}$
denote the diagonal matrices with the components of $\boldsymbol{p}_{1}$
and $\boldsymbol{p}_{2}$ as their diagonal elements, respectively,
then $X-\Gamma_{1}\geq0$, $\Gamma_{1}\geq0$, $X-\Gamma_{2}\geq0$,
$\Gamma_{2}\geq0$. Let $\epsilon$ be an arbitrary real number between
$0$ and $1$ and\begin{equation}
\boldsymbol{p}_{\epsilon}=\epsilon\boldsymbol{p}_{1}+\left(1-\epsilon\right)\boldsymbol{p}_{2},\label{eq:-20}\end{equation}
\begin{equation}
\Gamma_{\epsilon}=\epsilon\Gamma_{1}+\left(1-\epsilon\right)\Gamma_{2}.\label{eq:-21}\end{equation}
Since the sum of two positive matrices is still a positive matrix, we
have\begin{equation}
X-\Gamma_{\epsilon}=X-(\epsilon\Gamma_{1}+\left(1-\epsilon\right)\Gamma_{2})=\epsilon\left(X-\Gamma_{1}\right)+\left(1-\epsilon\right)\left(X-\Gamma_{2}\right)\geq0,\label{eq:-22}\end{equation}
\begin{equation}
\Gamma_{\epsilon}=\epsilon\Gamma_{1}+\left(1-\epsilon\right)\Gamma_{2}\geq0,\label{eq:-23}\end{equation}
which means that $\boldsymbol{p}_{\epsilon}\in\mathcal{S}$. Thus
$\mathcal{S}$ is a convex set. $\blacksquare$

Before deriving more properties of the optimum solution for unambiguous
discrimination, we introduce some notations and nomenclatures that
will be used later on here.

Let $\sigma_{1},\cdots,\sigma_{n}$ denote the eigenvalues of $X-\Gamma$
in decreasing order such that $\sigma_{1}\geq\cdots\geq\sigma_{n}$,
then the minimum eigenvalue $\sigma_{n}$ must satisfy $\sigma_{n}\geq0$
due to the positivity of $X-\Gamma$ according to Theorem \ref{positivity}. Since
$X-\Gamma$ depends on the parameters $\left\{ p_{1},\cdots,p_{n}\right\} $,
which is denoted by a real vector $\boldsymbol{p}$ for short, we
have $\sigma_{i}=\sigma_{i}(\boldsymbol{p})$, ($i=1,\cdots,n$).

For the convenience of description, the following notations and nomenclatures
will be used throughout this article.
\begin{itemize}
\item a \emph{point}: a vector $\boldsymbol{p}$ in $\mathbb{R}^{n}$;
\item the \emph{feasible set}: the set $\mathcal{S}$ defined in Theorem \ref{positivity};
\item the \emph{critical feasible region} (denoted by $\mathcal{R}_{CF}$):
the set of points $\boldsymbol{p}$ in $\mathcal{S}$ satisfying $\sigma_{n}\left(\boldsymbol{p}\right)=0$
and $\Gamma\geq0$;
\item the \emph{boundary} of the critical feasible region (denoted by $\mathcal{B}_{R}$):
the set of points $\boldsymbol{p}$ in the critical feasible region
$\mathcal{R}_{CF}$ with at least one zero component, i.e., $p_{i}=0$
for at least one $i\in\left\{ 1,\cdots,n\right\} $;
\item the \emph{interior part} of the critical feasible region (denoted
by $\Omega_{R}$): the set of points $\boldsymbol{p}$ in the critical
feasible region $\mathcal{R}_{CF}$ with $p_{i}>0$ for all $i\in\left\{ 1,\cdots,n\right\} $, i.e. $\Omega_{R}=\mathcal{R}_{CF}\backslash\mathcal{B}_{R}$ where "$\backslash$" denotes the set exclusion operator;
\item an \emph{interior point}: a point in the interior part $\Omega_{R}$
of the critical feasible region (but not off the critical feasible region);
\item a \emph{boundary point}: a point on the boundary $\mathcal{B}_{R}$
of the critical feasible region;
\item a \emph{singular point}: a point in the critical feasible region where
$\nabla\sigma_{n}\left(\boldsymbol{p}\right)$ does not exist (i.e.
$\sigma_{n}\left(\boldsymbol{p}\right)$ is degenerate) or $\nabla\sigma_{n}\left(\boldsymbol{p}\right)=\boldsymbol{0}$. Here $\nabla$ denotes the gradient operator.
\end{itemize}

It is clear that $\mathcal{B}_{R}\cup\Omega_{R}=\mathcal{R}_{CF}$, and $\mathcal{R}_{CF}\subset \mathcal{S}$.

We give an example of three pure states below to explain the nomenclatures defined above intuitively by graphics.
Suppose the states to be discriminated are\begin{equation}
|\psi_{1}\rangle=\left(1,0,0\right)^{T},\:|\psi_{2}\rangle=\frac{1}{\sqrt{5}}\left(1,2,0\right)^{T},\:|\psi_{3}\rangle=\frac{2}{\sqrt{17}}\Bigl(1,1,\frac{3}{2}\Bigr)^{T}.\label{eq:66}\end{equation}

\begin{figure}
\includegraphics[scale=0.5]{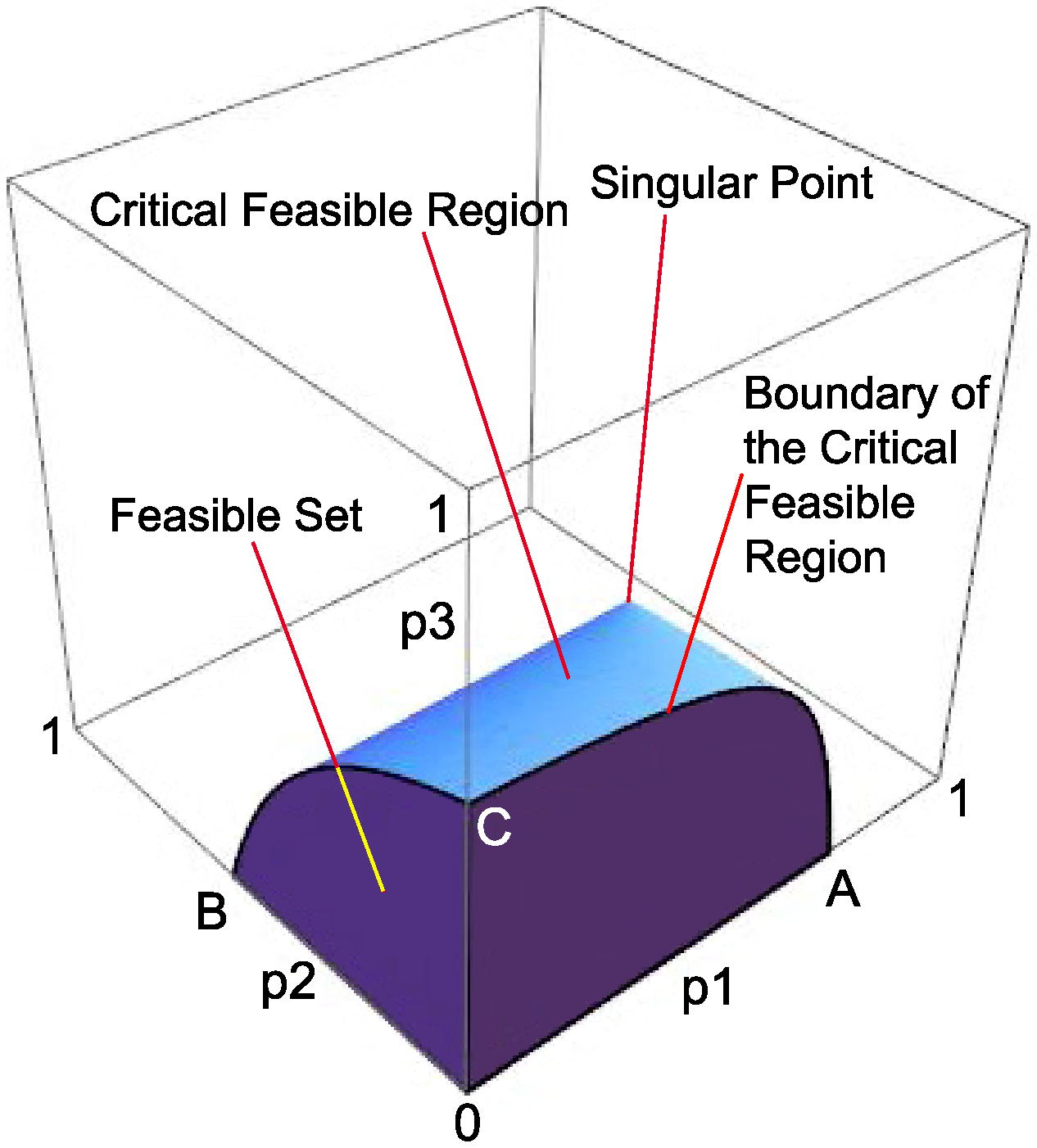}

\caption{\label{fig:General}(Color online)Feasible Set, critical feasible region, boundary and singular point. The whole shaded volume (including its surface) is the feasible set $\mathcal{S}$,  the upper curved surface of the feasible set is the critical feasible region $\mathcal{R}_{CF}$, and the black curved boundary line $\stackrel {\displaystyle{\frown}} {BC} \cup \stackrel {\displaystyle{\frown}} {AC} \cup\stackrel {\displaystyle{\frown}} {AB}$ ($\stackrel {\displaystyle{\frown}} {AB}$ is hidden and invisible in the figure) is the boundary $\mathcal{B}_{R}$ of the critical feasible region. The singular point is the point at which the curved surface is not smooth.
}

\end{figure}

Fig. \ref{fig:General} shows the feasible set $\mathcal{S}$, the
critical feasible region $\mathcal{R}_{CF}$, the boundary $\mathcal{B}_{R}$ of the
critical feasible region and the singular point
$\boldsymbol{p}_{S}$ for these three states. But it should be noted that the singular point does not always exist for every set of states. A necessary condition for the existence of a singular point in the case of three states is given in Case 3 of Sec. \ref{sub:Special-Cases}.
In addition, Fig.
\ref{fig:General} explicitly shows that the feasible set is convex,
which agrees with Theorem \ref{positivity}.

Now we give two lemmas that will be used in the proofs of the next several theorems.

\refstepcounter{lem}\label{convex_maximum}\emph{Lemma \thelem.} If there exists a point $\boldsymbol{p}_{0}$ in the
critical feasible region $\mathcal{R}_{CF}$ satisfying that $\boldsymbol{\gamma}\cdot\boldsymbol{p}\leq\boldsymbol{\gamma}\cdot\boldsymbol{p}_{0}$
for any point $\boldsymbol{p}$ in a sufficiently small neighborhood
$\Delta$ of $\boldsymbol{p}_{0}$, $\Delta\subset\mathcal{S}$,
then $\boldsymbol{\gamma}\cdot\boldsymbol{p}\leq\boldsymbol{\gamma}\cdot\boldsymbol{p}_{0}$
holds for any point $\boldsymbol{p}\in\mathcal{S}$.

\emph{Proof.} By contradiction. Suppose there exists such a point
$\boldsymbol{p}_{1}\in\mathcal{S}$ that $\boldsymbol{\gamma}\cdot\boldsymbol{p}_{1}>\boldsymbol{\gamma}\cdot\boldsymbol{p}_{0}$, let $\epsilon$ be a sufficiently small positive real number satisfying $0<\epsilon<1$, and let\begin{equation}
\boldsymbol{p}_{\epsilon}=\epsilon\boldsymbol{p}_{1}+\left(1-\epsilon\right)\boldsymbol{p}_{0},\label{eq:67}\end{equation}
 then $\boldsymbol{p}_{\epsilon}\in\mathcal{S}$ because of the convexity
of $\mathcal{S}$ proved in Theorem \ref{positivity}, and $\boldsymbol{p}_{\epsilon}\in\Delta$
since $\epsilon$ is sufficiently small. However,\begin{equation}
\boldsymbol{\gamma}\cdot\boldsymbol{p}_{\epsilon}=\boldsymbol{\gamma}\cdot\left(\epsilon\boldsymbol{p}_{1}+\left(1-\epsilon\right)\boldsymbol{p}_{0}\right)=\epsilon\boldsymbol{\gamma}\cdot\boldsymbol{p}_{1}+\left(1-\epsilon\right)\boldsymbol{\gamma}\cdot\boldsymbol{p}_{0}>\boldsymbol{\gamma}\cdot\boldsymbol{p}_{0}\label{eq:74}\end{equation}
which contradicts the assumption of this lemma. Therefore, $\boldsymbol{\gamma}\cdot\boldsymbol{p}\leq\boldsymbol{\gamma}\cdot\boldsymbol{p}_{0}$
holds for any point $\boldsymbol{p}$ in $\mathcal{S}$.$\blacksquare$

Intuitively, Lemma \ref{convex_maximum} tells us that if a linear function acting on a convex set  achieves a local maximal value at some point of the set, then that local maximal point must be the global maximum point that the function can reach over the whole set.

\refstepcounter{lem}\label{parallel}\emph{Lemma \thelem.} Suppose $\boldsymbol{a}$ and $\boldsymbol{b}$
are two non-zero real vectors in the $\mathbb{R}^{n}$ space. $\left(\boldsymbol{a}\cdot\boldsymbol{x}\right)\left(\boldsymbol{b}\cdot\boldsymbol{x}\right)\leq0$
for any vector $\boldsymbol{x}$, if and only if $\boldsymbol{a}$
and $\boldsymbol{b}$ are anti-parallel, i.e., $\boldsymbol{a}=-\lambda\boldsymbol{b}$
where $\lambda$ is positive factor.

\emph{Proof.} Necessity: By contradiction. Let $\boldsymbol{x}=\boldsymbol{a}_{0}+\boldsymbol{b}_{0}$,
where $\boldsymbol{a}_{0}=\frac{\boldsymbol{a}}{|\boldsymbol{a}|}$
and $\boldsymbol{b}_{0}=\frac{\boldsymbol{b}}{|\boldsymbol{b}|}$.
Let $\theta$ denote the angular between $\boldsymbol{a}$ and $\boldsymbol{b}$,
i.e., $\boldsymbol{a}\cdot\boldsymbol{b}=|\boldsymbol{a}||\boldsymbol{b}|\cos\theta$.
If $\boldsymbol{a}$ and $\boldsymbol{b}$ are not anti-parallel,
then $0\leq\theta<\pi$ by the well known \emph{Cauchy-Schwartz inequality}
\cite{cauchy schwartz} and $\boldsymbol{x}\neq\boldsymbol{0}$. Then
we have $\left(\boldsymbol{a}\cdot\boldsymbol{x}\right)\left(\boldsymbol{b}\cdot\boldsymbol{x}\right)=|\boldsymbol{a}||\boldsymbol{b}|\left(1+\cos\theta\right)^{2}>0$,
contradictory to the assumption.

Sufficiency: If $\boldsymbol{a}$ and $\boldsymbol{b}$ are anti-parallel,
i.e., $\boldsymbol{a}=-\lambda\boldsymbol{b}$ ($\lambda>0$), then for any non-zero
$\boldsymbol{x}$, $\left(\boldsymbol{a}\cdot\boldsymbol{x}\right)\left(\boldsymbol{b}\cdot\boldsymbol{x}\right)=-\lambda\left(\boldsymbol{b}\cdot\boldsymbol{x}\right)^{2}\leq0$.
$\blacksquare$

Now we are going to show some important properties of the optimum point, from which we will establish two sets of new equations that the optimum solution has to satisfy later in different situations.

\refstepcounter{thm}\label{mineigen}\emph{Theorem \thethm (Minimum Eigenvalue).} i) The optimum point $\boldsymbol{p}_{opt}$
must be a point in the critical feasible region $\mathcal{R}_{CF}$,
i.e.,\begin{equation}
\sigma_{n}\left(\boldsymbol{p}_{opt}\right)=0.\label{eq:-19}\end{equation}
ii) If the optimum point $\boldsymbol{p}_{opt}$ is a non-singular
point in the interior part $\Omega_{R}$ of the critical
feasible region, $\boldsymbol{p}_{opt}\in \Omega_{R}$,
then\begin{equation}
\nabla\sigma_{n}\left(\boldsymbol{p}\right)|_{\boldsymbol{p}_{opt}}=-\boldsymbol{\gamma}.\label{eq:-3}\end{equation}
iii) Conversely, if there exists a point $\boldsymbol{p}_{0}$ satisfying
$\nabla\sigma_{n}\left(\boldsymbol{p}\right)|_{\boldsymbol{p}_{0}}=-\boldsymbol{\gamma}$
in the critical feasible region $\mathcal{R}_{CF}$, $\boldsymbol{p}_{0}$
must be the global optimum point.

\emph{Proof.} i) Since $\sigma_{n}$ is the minimum eigenvalue of the matrix $X-\Gamma$ and $X-\Gamma$ is positive as proved in Theorem \ref{positivity}, we have $\sigma_{n}\geq0$. Thus, in order to prove Eq. \eqref{eq:-19}, we only need to show that $\sigma_{n}$ cannot be greater than zero at the optimum point. We prove it by contradiction.

If $\sigma_{n}\left(\boldsymbol{p}_{opt}\right)>0$, we let $\boldsymbol{p}_{opt}$ be changed a little
by $\delta\boldsymbol{p}$, then we can always
find such $\delta\boldsymbol{p}$ that $\boldsymbol{\gamma}\cdot\delta\boldsymbol{p}>0$
and at the same time $\sigma_{n}\left(\boldsymbol{p}_{opt}+\delta\boldsymbol{p}\right)\geq0$,
$\Gamma+\delta\Gamma\geq0$, where $\delta\Gamma$ is a diagonal matrix
with $\delta p_{1},\cdots,\delta p_{n}$ as its diagonal elements
(such $\delta\boldsymbol{p}$ always exists as we can construct it
simply by taking $\delta p_i>0$ and $\delta p_i$ sufficiently small, for all $i=1,\cdots,n$).
This implies that the point $\boldsymbol{p}_{opt}+\delta\boldsymbol{p}$ in the feasible set $\mathcal{S}$ satisfies $\boldsymbol{\gamma}\cdot(\boldsymbol{p}_{opt}+\delta\boldsymbol{p})>\boldsymbol{\gamma}\cdot\boldsymbol{p}_{opt}$, which contradicts the assumption that $\boldsymbol{\gamma}\cdot\boldsymbol{p}$
reaches the maximum at $\boldsymbol{p}_{opt}$, so Eq. \eqref{eq:-19}
holds.

It seems that the above proof may be applied to other eigenvalues of $X-\Gamma$ in a similar way, but if any eigenvalue other than the minimal one is equal to zero, then the minimal eigenvalue would be negative and $X-\Gamma$ would not keep positive, violating Eq. \eqref{eq:1}.

ii) Since $\boldsymbol{p}_{opt}$ is an interior optimum point in
the critical feasible region $\mathcal{R}_{CF}$, if we change $\boldsymbol{p}_{opt}$
a little by any $\delta\boldsymbol{p}$ such that $\sigma_{n}(\boldsymbol{p}_{opt}+\delta\boldsymbol{p})\geq0$,
it must be that $\boldsymbol{\gamma}\cdot\delta\boldsymbol{p}\leq0$,
otherwise $\boldsymbol{\gamma}\cdot\boldsymbol{p}$ is not maximal
at $\boldsymbol{p}_{opt}$. Considering $\sigma_{n}\left(\boldsymbol{p}_{opt}\right)=0$,
the inequality $\sigma_{n}(\boldsymbol{p}_{opt}+\delta\boldsymbol{p})\geq0$ can be converted to\begin{equation}
\delta\sigma_{n}\left(\boldsymbol{p}\right)|_{\boldsymbol{p}_{opt}}\geq0.\label{eq:-15}\end{equation}
Since $\boldsymbol{p}_{opt}$ is not a singular point, $\nabla\sigma_{n}\left(\boldsymbol{p}\right)|_{\boldsymbol{p}_{opt}}$
exists, so\begin{equation}
\delta\sigma_{n}\left(\boldsymbol{p}\right)=\sigma_{n}\left(\boldsymbol{p}_{opt}+\delta\boldsymbol{p}\right)-\sigma_{n}\left(\boldsymbol{p}_{opt}\right)
=\nabla\sigma_{n}\left(\boldsymbol{p}\right)|_{\boldsymbol{p}_{opt}}\cdot\delta\boldsymbol{p}.\label{eq:-16}\end{equation}
Substituting Eq. \eqref{eq:-16} into \eqref{eq:-15}, we have\begin{equation}
\nabla\sigma_{n}\left(\boldsymbol{p}\right)|_{\boldsymbol{p}_{opt}}\cdot\delta\boldsymbol{p}\geq0.\label{eq:-2}\end{equation}
Comparing Eq. \eqref{eq:-2} with $\boldsymbol{\gamma}\cdot\delta\boldsymbol{p}\leq0$,
it follows that $\boldsymbol{\gamma}$ and $\nabla\sigma_{n}\left(\boldsymbol{p}\right)|_{\boldsymbol{p}_{opt}}$
must be anti-parallel according to Lemma \ref{parallel},
so\begin{equation}
\nabla\sigma_{n}\left(\boldsymbol{p}\right)|_{\boldsymbol{p}_{opt}}=-\alpha\boldsymbol{\gamma},\;\alpha>0.\label{eq:-5}\end{equation}
We will prove that $\alpha=1$ in the proof of Theorem \ref{nonsigular}, thus
Eq. \eqref{eq:-3} holds.

iii) If there exists a point $\boldsymbol{p}_{0}$ where $\nabla\sigma_{n}\left(\boldsymbol{p}\right)|_{\boldsymbol{p}_{0}}=-\boldsymbol{\gamma}$
in the critical feasible region $\mathcal{R}_{CF}$, any small $\delta\boldsymbol{p}$
such that $\sigma_{n}\left(\boldsymbol{p}_{0}+\delta\boldsymbol{p}\right)=\delta\sigma_{n}\left(\boldsymbol{p}\right)=\nabla\sigma_{n}\left(\boldsymbol{p}\right)|_{\boldsymbol{p}_{0}}\cdot\delta\boldsymbol{p}\geq0$
will lead to $\boldsymbol{\gamma}\cdot\delta\boldsymbol{p}=-\nabla\sigma_{n}\left(\boldsymbol{p}\right)|_{\boldsymbol{p}_{0}}\cdot\delta\boldsymbol{p}\leq0$,
so $\boldsymbol{\gamma}\cdot\boldsymbol{p}$ reaches maximal at $\boldsymbol{p}_{0}$
in a small neighborhood of $\boldsymbol{p}_{0}$, and according to
Lemma \ref{convex_maximum}, $\boldsymbol{p}_{0}$ must be the global optimum point where
$\boldsymbol{\gamma}\cdot\boldsymbol{p}$ reaches the global maximum.
$\blacksquare$

It should be pointed out that a result equivalent to our result i)
in Theorem \ref{mineigen} was obtained in Ref. \cite{chefles unambiguous 1}, where
the maximum eigenvalue of $\widetilde{\Phi}\Gamma\widetilde{\Phi}^{\dagger}$
is proved to be exactly $1$ when the unambiguous state discrimination
scheme is optimum. Besides, Eq. \eqref{eq:-19} implies that\begin{equation}
\det\left(X-\Gamma\right)=0\label{eq:2}\end{equation}
at the optimum point, and a physical interpretation of this fact was given
in \cite{determinant 0}.

To prove the next theorem, we first give another lemma \cite{matrix analysys}.

\refstepcounter{lem}\label{eigenvalue_sum}\emph{Lemma \thelem.} If $\tau_{1},\cdots,\tau_{n}$ are the eigenvalues
of an $n\times n$ matrix $A$, then $\underset{1\leq i_{1}<\cdots<i_{k}\leq n}{\sum}\tau_{i_{1}}\cdots\tau_{i_{k}}=\sum\left(\text{all $k\times k$ principal minors of $A$}\right)$.

Based on Theorem \ref{mineigen}, we now present the following theorem which can
be conveniently used to obtain the maximum average success probability
and the optimum measurement strategy when the optimum point is not singular and in
the interior part $\Omega_{R}$ of the critical feasible region.

\refstepcounter{thm}\label{nonsigular}\emph{Theorem \thethm (Interior Non-Singular Solution).} Let $M_{k}\left(\boldsymbol{p}\right)$ denote the principle minor of order $n-1$ associated with the $k$th diagonal element of $X-\Gamma$, then a non-singular point $\boldsymbol{p}_{0}$ in the interior part $\Omega_{R}$ of the critical feasible region is the optimum point if and only if $\boldsymbol{p}_{0}$
is a solution of the following set of equations\begin{equation}
\begin{cases}
M_{1}\left(\boldsymbol{p}\right)=\gamma_{1}\lambda\\
\quad\vdots\\
M_{n}\left(\boldsymbol{p}\right)=\gamma_{n}\lambda\\
\det\left(X-\Gamma\right)=0\end{cases},\label{eq:5}\end{equation}
for some positive number $\lambda>0$, and satisfies the positivity condition \eqref{eq:1} at the same time.

\emph{Proof.} First, we prove the "only if" part of the theorem. Let's consider the variation
of $\det\left(X-\Gamma\right)$. Since $\det\left(X-\Gamma\right)=\sigma_{1}\dotsm\sigma_{n}$, we have\begin{equation}
\delta\det\left(X-\Gamma\right)=\left(\delta\sigma_{1}\right)\sigma_{2}\dotsm\sigma_{n}+\sigma_{1}\left(\delta\sigma_{2}\right)\sigma_{3}\dotsm\sigma_{n}+\cdots+\sigma_{1}\sigma_{2}\dotsm\sigma_{n-1}\delta\sigma_{n}.\label{eq:-6}\end{equation}
If $\boldsymbol{p}_{0}$ is the optimum point that achieves the maximum average probability, then
$\sigma_{n}\left(\boldsymbol{p}_{0}\right)=0$ according to Theorem \ref{mineigen},
Eq. \eqref{eq:-6} can be simplified to\begin{equation}
\delta\det\left(X-\Gamma\right)|_{\boldsymbol{p}_{0}}=\left(\sigma_{1}\sigma_{2}\dotsm\sigma_{n-1}\delta\sigma_{n}\right)|_{\boldsymbol{p}_{0}}.\label{eq:-12}\end{equation}
Since $\boldsymbol{p}_{0}$ is not a singular point, $\nabla\sigma_{n}\left(\boldsymbol{p}\right)|_{\boldsymbol{p}_{0}}$
exists and $\delta\sigma_{n}\left(\boldsymbol{p}\right)|_{\boldsymbol{p}_{0}}=\nabla\sigma_{n}\left(\boldsymbol{p}\right)|_{\boldsymbol{p}_{0}}\cdot\delta\boldsymbol{p}$.
From $\delta\det\left(X-\Gamma\right)=\nabla\det\left(X-\Gamma\right)\cdot\delta\boldsymbol{p}$
and Eq. \eqref{eq:-12}, we have\begin{equation}
\nabla\det\left(X-\Gamma\right)|_{\boldsymbol{p}_{0}}=\sigma_{1}\sigma_{2}\dotsm\sigma_{n-1}\nabla\sigma_{n}\left(\boldsymbol{p}\right)|_{\boldsymbol{p}_{0}}.\label{eq:-14}\end{equation}
Again using $\sigma_{n}\left(\boldsymbol{p}_{0}\right)=0$, we can get\begin{equation}
\nabla\det\left(X-\Gamma\right)|_{\boldsymbol{p}_{0}}=\left(\sigma_{1}\dotsm\sigma_{n-1}+\sigma_{1}\dotsm\sigma_{n-2}\sigma_{n}+\cdots+\sigma_{2}\dotsm\sigma_{n}\right)\nabla\sigma_{n}\left(\boldsymbol{p}\right)|_{\boldsymbol{p}_{0}}.\label{eq:-13}\end{equation}
According to Lemma \ref{eigenvalue_sum}, we have\begin{equation}
\nabla\det\left(X-\Gamma\right)|_{\boldsymbol{p}_{0}}=\left(M_{1}\left(\boldsymbol{p}_{0}\right)+\cdots+M_{n}\left(\boldsymbol{p}_{0}\right)\right)\nabla\sigma_{n}\left(\boldsymbol{p}\right)|_{\boldsymbol{p}_{0}}.\label{eq:-4}\end{equation}

On the other hand,\begin{equation}
\nabla\det\left(X-\Gamma\right)=\left(\boldsymbol{e}_{1}\frac{\partial}{\partial p_{1}}+\cdots+\boldsymbol{e}_{n}\frac{\partial}{\partial p_{n}}\right)\det\left(X-\Gamma\right),\label{eq:60}\end{equation}
where $\boldsymbol{e}_{i}$ is the orthonormal basis vector in the
space $\mathbb{R}^{n}$ associated with the coordinate $p_{i}$. By
performing Laplace expansion on the determinant of $X-\Gamma$ along
the $k$th row (or column), we can see that\begin{equation}
\frac{\partial}{\partial p_{k}}\det\left(X-\Gamma\right)=-M_{k}\left(\boldsymbol{p}\right),\label{eq:51}\end{equation}
so\begin{equation}
\nabla\det\left(X-\Gamma\right)=-\left(M_{1}\left(\boldsymbol{p}\right)\boldsymbol{e}_{1}+\cdots+M_{n}\left(\boldsymbol{p}\right)\boldsymbol{e}_{n}\right).\label{eq:40}\end{equation}

Comparing Eq. \eqref{eq:-4} and \eqref{eq:40} and substituting Eq.
\eqref{eq:-5} into \eqref{eq:-4}, we have\begin{equation}
-\left(M_{1}\left(\boldsymbol{p}_{0}\right)\boldsymbol{e}_{1}+\cdots+M_{n}\left(\boldsymbol{p}_{0}\right)\boldsymbol{e}_{n}\right)=-\left(M_{1}\left(\boldsymbol{p}_{0}\right)+\cdots+M_{n}\left(\boldsymbol{p}_{0}\right)\right)\alpha\boldsymbol{\gamma},\label{eq:-24}\end{equation}
which results in\begin{equation}
M_{k}\left(\boldsymbol{p}_{0}\right)=\alpha\gamma_{k}\left(M_{1}\left(\boldsymbol{p}_{0}\right)+\cdots+M_{n}\left(\boldsymbol{p}_{0}\right)\right),\; k=1,\cdots,n.\label{eq:59}\end{equation}

Considering that $X-\Gamma\geq0$, we have $M_{k}\left(\boldsymbol{p}_{0}\right)\geq0$ for all $k=1,\cdots,n$. Since the theorem assumes that the optimum point is not singular, the zero eigenvalue is not degenerate and $\nabla\sigma_{n}\left(\boldsymbol{p}\right)|_{\boldsymbol{p}_{0}}\neq\boldsymbol{0}$,
then according to Eq. \eqref{eq:-14}, we have\begin{equation}\nabla\det\left(X-\Gamma\right)|_{\boldsymbol{p}_{0}}\neq\boldsymbol{0}.
\label{eq:bbb}\end{equation}So according to Eq. \eqref{eq:40} we can see that the $M_{k}\left(\boldsymbol{p}_{0}\right)$'s are not all zeros and\begin{equation}
M_{1}\left(\boldsymbol{p}_{0}\right)+\cdots+M_{n}\left(\boldsymbol{p}_{0}\right)>0.\label{eq:ccc}\end{equation}
By summing up Eq. \eqref{eq:59} for $i=1,\cdots,n$ and using $\gamma_{1}+\cdots\gamma_{n}=1$ and Eq. \eqref{eq:ccc}, we immediately have $\alpha=1$.
(This completes the remaining part of the proof for part ii) of Theorem \ref{mineigen}.)

Thus, Eq. \eqref{eq:59} can be simplified to\begin{equation}
M_{k}\left(\boldsymbol{p}_{0}\right)=\gamma_{k}\left(M_{1}\left(\boldsymbol{p}_{0}\right)+\cdots+M_{n}\left(\boldsymbol{p}_{0}\right)\right),\; k=1,\cdots,n,\label{eq:52}\end{equation}
which immediately leads to Eq. \eqref{eq:5}.
Considering Eq. \eqref{eq:ccc}, there must be $\lambda>0$ according to Eq. \eqref{eq:5}.

This concludes the proof of the "only if" part of the theorem.

Next, we prove the "if" part of the theorem. Considering $X-\Gamma\geq0$, $\det\left(X-\Gamma\right)=0$ in Eq. \eqref{eq:5} and that $\sigma_{n}\left(\boldsymbol{p}_{0}\right)$ is the minimal eigenvalue of $X-\Gamma$, it follows straightforwardly that
\begin{equation}
\sigma_{n}\left(\boldsymbol{p}_{0}\right)=0.\label{eq:aaa}\end{equation}
And since $\lambda>0$, $M_{k}\left(\boldsymbol{p}_{0}\right)>0$,
$k=1,\cdots,n$, thus the rank of $X-\Gamma$ is $n-1$ and $\nabla\det\left(X-\Gamma\right)|_{\boldsymbol{p}_{0}}\neq\boldsymbol{0}$, which imply that $\boldsymbol{p}_{0}$ is not a singular point.
It can be seen that Eq. \eqref{eq:-4} still holds here,
so substituting \eqref{eq:5} into Eqs. \eqref{eq:-4} and \eqref{eq:40},
one can have\begin{equation}
-\left(\gamma_{1}\lambda\boldsymbol{e}_{1}+\cdots+\gamma_{n}\lambda\boldsymbol{e}_{n}\right)=\lambda\nabla\sigma_{n}\left(\boldsymbol{p}\right)|_{\boldsymbol{p}_{0}},\label{eq:61}\end{equation}
where $\gamma_{1}+\cdots\gamma_{n}=1$ has been used. Therefore we
get $\nabla\sigma_{n}\left(\boldsymbol{p}\right)|_{\boldsymbol{p}_{0}}=-\boldsymbol{\gamma}$,
and together with \eqref{eq:aaa} it implies that $\boldsymbol{p}_{0}$
is the (global) optimum point according to part iii) of Theorem \ref{mineigen}. This concludes the proof of the "if" part of the theorem.
$\blacksquare$

Theorem \ref{mineigen} and Theorem \ref{nonsigular} describe the properties of the optimum point
when it is a non-singular point in the interior part $\Omega_{R}$
of the critical feasible region, and Theorem \ref{nonsigular} also gives a way to
find such an optimum point. When one obtains a solution from Eq. \eqref{eq:5} for some $\lambda>0$,
he or she has to verify whether it satisfies Eq. \eqref{eq:1}. If it does,
this solution is the optimum solution. However, if \eqref{eq:5}
has no solution satisfying $\lambda>0$ and \eqref{eq:1}, then the optimum point is either a point
on the boundary $\mathcal{B}_{R}$ or a singular point (if it exists)
of the critical feasible region $\mathcal{R}_{CF}$. The next theorem
is to characterize the properties of the optimum point when it is
on the boundary $\mathcal{B}_{R}$ and give a method to work out the
optimum point in that situation.

For the simplicity of later description, we further define some new
notations here. We denote the part of the critical feasible region
$\mathcal{R}_{CF}$ where $p_{i_{1}}=0,\cdots,\ p_{i_{k}}=0$ as $\mathcal{B}_{R}\left(i_{1},\cdots,i_{k}\right)$,
and any $\mathcal{B}_{R}\left(i_{1},\cdots,i_{k}\right)$ will be
called a $\left(n-k-1\right)$\emph{-dimensional boundary} of the
critical feasible region (since the critical feasible region itself is of
dimension $n-1$ in the $\mathbb{R}^{n}$ space). We again take the three state \eqref{eq:66} as an example. In Fig. \ref{fig:General}, the black boundary line excluding the points A, B and C is the 1-dimensional boundary of the critical feasible region, and the points A, B and C are the 0-dimensional boundaries of the critical feasible region.

\refstepcounter{thm}\label{boundary}\emph{Theorem \thethm  (Boundary Solution).} A point $\boldsymbol{p}_{0}$
on a $\left(n-k_{0}-1\right)$-dimensional boundary $\mathcal{B}_{R}\left(i_{1},\cdots,i_{k_{0}}\right)$
but not on any lower dimensional boundary, i.e., $p_{i_{1}}=0,\cdots,\ p_{i_{k_{0}}}=0$
and $p_{j}>0$, $\forall\: j\neq i_{1},\cdots,i_{k_{0}}$, is the optimum point
if and only if it is a solution of\begin{equation}
\begin{cases}
M_{i}\left(\boldsymbol{p}\right)|_{p_{i_{1}}=0,\cdots,p_{i_{k_{0}}}=0}=\gamma_{i}\lambda & \forall i\in\left\{ 1,\cdots,n\right\} \backslash\left\{ i_{1},\cdots,i_{k_{0}}\right\} \\
\det\left(X-\Gamma\right)|_{p_{i_{1}}=0,\cdots,p_{i_{k_{0}}}=0}=0\end{cases},\label{eq:78}\end{equation}
satisfying $\lambda>0$, the positivity constraints \eqref{eq:1}
and\begin{equation}
\begin{cases}
M_{i_{1}}\left(\boldsymbol{p}\right)|_{p_{i_{1}}=0,\cdots,p_{i_{k_{0}}}=0}\geq\lambda\gamma_{i_1}\\
\qquad\vdots\\
M_{i_{k_{0}}}\left(\boldsymbol{p}\right)|_{p_{i_{1}}=0,\cdots,p_{i_{k_{0}}}=0}\geq\lambda\gamma_{i_{k_{0}}}\end{cases}.\label{eq:79}\end{equation}

\emph{Proof.} First, we prove the "only if" part of the theorem. If $\boldsymbol{p}_{0}$ is an optimum point
on the $\left(n-k_{0}-1\right)$-dimensional boundary $\mathcal{B}_{R}\left(i_{1},\cdots,i_{k_{0}}\right)$
but not on any lower dimensional boundary, let $\boldsymbol{p}_{0}$
be changed to $\boldsymbol{p}_{0}+\delta\boldsymbol{p}$ on $\mathcal{B}_{R}\left(i_{1},\cdots,i_{k_{0}}\right)$
by any small $\delta\boldsymbol{p}$ satisfying $\delta p_{i_{1}}=0,\cdots,\delta p_{i_{k_{0}}}=0$,
then there must be
\begin{eqnarray}
\delta\det\left(X-\Gamma\right)|_{\boldsymbol{p}_{0}} & = & \nabla\det\left(X-\Gamma\right)|_{\boldsymbol{p}_{0}}\cdot\delta\boldsymbol{p}
=-\sum_{i\neq i_{1},\cdots,i_{k_{0}}}M_{i}\left(\boldsymbol{p}_{0}\right)\delta p_{i}=0,\label{eq:32}
\end{eqnarray}
where we have used Eq. \eqref{eq:40}, and\begin{equation}
\boldsymbol{\gamma}\cdot\delta\boldsymbol{p}=\sum_{i\neq i_{1},\cdots,i_{k_{0}}}\gamma_{i}\delta p_{i}\leq0.\label{eq:81}\end{equation}
So\begin{equation}
\left(M_{1}\left(\boldsymbol{p}_{0}\right),\cdots,M_{i}\left(\boldsymbol{p}_{0}\right),\cdots,M_{n}\left(\boldsymbol{p}_{0}\right)\right)|_{i\neq i_{1},\cdots,i_{k_{0}}}\propto\left(\gamma_{1},\cdots,\gamma_{i},\cdots,\gamma_{n}\right)|_{i\neq i_{1},\cdots,i_{k_{0}}}\label{eq:3}\end{equation}
according to Lemma \ref{parallel}, therefore, Eq. \eqref{eq:78} holds.

On the other hand, let $\boldsymbol{p}_{0}$ be changed to $\boldsymbol{p}_{0}+\delta\boldsymbol{p}^{\prime}$
in the critical feasible region $\mathcal{R}_{CF}$ by another arbitrary
small $\delta\boldsymbol{p}^{\prime}$ satisfying that $\delta p_{i_{1}}^{\prime}\geq0,\cdots,\delta p_{i_{k_{0}}}^{\prime}\geq0$,
then there must be\begin{equation}\label{eq:82}\begin{split}
\delta\det\left(X-\Gamma\right)|_{\boldsymbol{p}_{0}} & =-\sum_{i=1}^{n}M_{i}\left(\boldsymbol{p}_{0}\right)\delta p_{i}^{\prime}
\\
 & =-\sum_{i=i_{1},\cdots,i_{k_{0}}}M_{i}\left(\boldsymbol{p}_{0}\right)\delta p_{i}^{\prime}-\sum_{i\neq i_{1},\cdots,i_{k_{0}}}M_{i}\left(\boldsymbol{p}_{0}\right)\delta p_{i}^{\prime}
 \\
 & =-\sum_{i=i_{1},\cdots,i_{k_{0}}}M_{i}\left(\boldsymbol{p}_{0}\right)\delta p_{i}^{\prime}-\sum_{i\neq i_{1},\cdots,i_{k_{0}}}\lambda\gamma_{i}\delta p_{i}^{\prime}=0\end{split}\end{equation}
where we have used Eq. \eqref{eq:78}, and\begin{equation}
\boldsymbol{\gamma}\cdot\delta\boldsymbol{p}^{\prime}=\sum_{i=1}^{n}\gamma_{i}\delta p_{i}^{\prime}=\sum_{i=i_{1},\cdots,i_{k_{0}}}\gamma_{i}\delta p_{i}^{\prime}+\sum_{i\neq i_{1},\cdots,i_{k_{0}}}\gamma_{i}\delta p_{i}^{\prime}\leq0.\label{eq:83}\end{equation}
Substituting Eq. \eqref{eq:82} into \eqref{eq:83}, we have\begin{equation}
\boldsymbol{\gamma}\cdot\delta\boldsymbol{p}^{\prime}=\sum_{i=1}^{n}\gamma_{i}\delta p_{i}^{\prime}=\sum_{i=i_{1},\cdots,i_{k_{0}}}\left(\gamma_{i}-\frac{M_{i}\left(\boldsymbol{p}_{0}\right)}{\lambda}\right)\delta p_{i}^{\prime}\leq0.\label{eq:84}\end{equation}

Considering $\delta p_{i_{1}}^{\prime}\geq0,\cdots,\delta p_{i_{k_{0}}}^{\prime}\geq0$
and $\lambda>0$, it can been seen that Eq. \eqref{eq:79} holds according
to Eq. \eqref{eq:84}. This concludes the proof of the "only if" part of the theorem.

The "if" part of the theorem can be directly proved by reversing the above reasoning
and using Lemma \ref{convex_maximum}, so we are not going to show the details here.
$\blacksquare$

Since the feasible set $\mathcal{S}$ is a closed convex set confined
in a finite region of $\mathbb{R}^{n}$, the optimum point where the
average success probability $\mbox{\ensuremath{\boldsymbol{\gamma}\cdot}}\boldsymbol{p}$
reaches the maximum always exists. Thus if Eq. \eqref{eq:78} for
any $\left(n-k_{0}-1\right)$-dimensional boundary ($1\leq k_{0}\leq n-1$)
does not have a solution satisfying $\lambda>0$, the positivity constraints
\eqref{eq:1} and Eq. \eqref{eq:79} while the optimum point is also
not an interior non-singular point in the critical feasible region
$\mathcal{R}_{CF}$, the optimum point can only be a singular point
then. We know that a singular
point is a point in the critical feasible region where $\sigma_{n}\left(\boldsymbol{p}\right)=0$
is degenerate or $\nabla\sigma_{n}\left(\boldsymbol{p}\right)=\boldsymbol{0}$,
so when the optimum point is singular,
$\nabla\det\left(X-\Gamma\right)|_{\boldsymbol{p}_{0}}=\boldsymbol{0}$
according to Eq. \eqref{eq:-14}, which, together with \eqref{eq:40},
implies that all $M_{k}\left(\boldsymbol{p}\right)=0$ for all $k=1,\cdots,n$.
Thus a singular point can be obtained as a solution of \eqref{eq:5} with $\lambda=0$ and the positivity conditions \eqref{eq:1}.

\refstepcounter{rmk}\label{remark:singular}\emph{Remark \thermk.} It should be pointed out that given a set of linearly independent states, if there exists a singular point in the critical feasible
region $\mathcal{R}_{CF}$, then that singular point could be the optimum
point for a range of different $\boldsymbol{\gamma}$'s, since the
normal vector of the critical feasible region $\mathcal{R}_{CF}$
changes discontinuously in the neighborhood of a singular point.

In this subsection we have mainly studied the properties of the optimum
point and obtained the equations that the optimum point must satisfy
in different situations. We summarize our method to find the optimum
point as follows:

\emph{Step I.} Try to solve Eq. \eqref{eq:5} in Theorem \ref{nonsigular} and see
whether there exists a solution satisfying the positivity constraints
$\lambda>0$ and Eq. \eqref{eq:1}. If such a solution exists, it
is exactly the optimum point we try to find.

\emph{Step II.} If the set of equations \eqref{eq:5} does not have
a solution that satisfies $\lambda>0$ and Eq. \eqref{eq:1}, one
has to continue to search for the optimum point on the boundary $\mathcal{B}_{R}$
of the critical feasible region using Theorem \ref{boundary}. In detail, one can
first solve Eq. \eqref{eq:78} on all $\left(n-2\right)$-dimensional
boundaries and see whether there exists a solution that satisfies
$\lambda>0$, the positivity constraints \eqref{eq:1} and Eq. \eqref{eq:79}.
If such a solution exists, it is exactly the optimum point; otherwise,
one should further search on all $\left(n-3\right)$-dimensional boundaries,
$\left(n-4\right)$-dimensional boundaries, \ldots{}, until such
a solution is found or all boundaries of dimension lower than $n-1$
have been searched. If such a solution is found, it is exactly the
optimum solution.

\emph{Step III.} If the optimum point is not found in the above two
steps, then it must be a singular point and can be obtained by solving
Eq. \eqref{eq:5} with $\lambda=0$. If there exists more than one
singular point on the critical feasible region $\mathcal{R}_{CF}$,
the one that maximizes $\mbox{\ensuremath{\boldsymbol{\gamma}\cdot}}\boldsymbol{p}$
is the optimum point.

By Theorem \ref{nonsigular} and \ref{boundary}, we give two sets of explicit analytical equations for solving the unambiguous discrimination problem in different situations. We can use them to work out analytical solutions or obtain some analytical relations for the problem (see examples in Sec. \ref{sec:Maximum-Average-Success}, \ref{sec:Generalization-of-Equal-Probability} and \ref{sec:Special-Case:}). However, since Eqs. \eqref{eq:5}
and \eqref{eq:78} are nonlinear and the variables $p_1,\cdots,p_n$ are tightly coupled in the equations, maybe only numeric solutions can be obtained for these equations in some situations. A lot of sophisticated numerical techniques like Newton's method (including many of its variants), hybrid Krylov methods and so on have been developed to solve such nonlinear equations \cite{numerical 1,numerical 2,numerical 3,numerical 4}.

It is worth mentioning that other methods such as semidefinite programming \cite{semidefi 0,semidefi 1,ieee epm} have been developed to solve this unambiguous discrimination problem. Those methods are  developed from some classical numerical analysis theories and they are mostly suitable for finding numeric solutions, while our method is developed purely by algebra and go in a totally different way, aiming at providing a new tool to treat the problem analytically.

\subsection{Geometrical view and numerical example\label{sub:Numerical-Example}}

A geometrical method is given in \cite{peres unambiguous 1998} to
solve the optimum unambiguous discrimination problem for three pure
states, mainly for the situation when the optimum point is an interior
non-singular point. In this subsection, we are going to give a similar
but more complete geometrical way to illustrate the problem and the
results we obtain in the previous subsection, and we calculate a numerical
example illustrated with corresponding graphics to explicitly show
the geometrical meanings.

Geometrically, $\overline{p}=\boldsymbol{\gamma}\cdot\boldsymbol{p}=\gamma_{1}p_{1}+\cdots+\gamma_{n}p_{n}$
can be perceived as an $\left(n-1\right)$-dimensional plane in the
$\mathbb{R}^{n}$ space, and the critical feasible region $\sigma_{n}\left(\boldsymbol{p}\right)=0$
can be perceived as a curved surface in $\mathbb{R}^{n}$.

It can be shown easily that the vertical distance from the origin
of the coordinate system (with $p_{1},\cdots,p_{n}$ as the coordinates)
to the plane $\bar{p}=\boldsymbol{\gamma}\cdot\boldsymbol{p}$ is\begin{equation}
\frac{\bar{p}}{\sqrt{\gamma_{1}^{2}+\cdots+\gamma_{n}^{2}}},\label{eq:62}\end{equation}
so the average success probability $\bar{p}$ characterizes the vertical distance
between the origin and the plane $\bar{p}=\boldsymbol{\gamma}\cdot\boldsymbol{p}$
in a geometrical view. Therefore, the problem of optimum unambiguous
discrimination of pure states can be translated to the problem of
finding an optimum point in the feasible set $\mathcal{S}$ at
which the plane with fixed normal vector $\left(\gamma_{1},\gamma_{2},\gamma_{3}\right)$
(unnormalized) is most distant from the origin. Obviously the optimum
point must lie in the critical feasible region $\mathcal{R}_{CF}$, which is the "surface" of the feasible set $\mathcal{S}$,
and this is in accordance with i) of Theorem \ref{mineigen}.

Now suppose the plane $\overline{p}=\boldsymbol{\gamma}\cdot\boldsymbol{p}$
is moved by parallel shifts, i.e., by changing $\bar{p}$ while the normal vector
$\left(\gamma_{1},\gamma_{2},\gamma_{3}\right)$ keeps fixed. If the
plane can be tangent with the critical feasible region $\mathcal{R}_{CF}$
when $\bar{p}$ is equal to some $\overline{p}_{0}$, the distance
from the origin to the plane is then maximized and the tangent point
is exactly the optimum point $\boldsymbol{p}_{opt}$, and $\bar{p}_{0}$
is the maximum average success probability.

When the plane and the critical feasible region $\mathcal{R}_{CF}$
are tangent, their normal vectors at the tangent point should be parallel
or anti-parallel. This implies that $\nabla\left(\boldsymbol{\gamma}\cdot\boldsymbol{p}-\bar{p}\right)=\zeta\nabla\det\left(X-\Gamma\right)$,
where $\zeta\in\mathbb{R}$ and $\zeta\neq0$. Using Eq. \eqref{eq:40},
this equation can be simplified to Eq. \eqref{eq:5}.

However, if the plane can never
be tangent with the critical feasible region when the plane is moved by any parallel shift,
it means that a non-singular optimum point does not exist in
the interior part of the critical feasible region, then the optimum
point is either a boundary point on $\mathcal{B}_{R}$ or a singular
point in the critical feasible region. And if the optimum point is
on some $\left(n-k-1\right)$-dimensional boundary, the plane is then
tangent with that $\left(n-k-1\right)$-dimensional boundary, resulting
in Eq. \eqref{eq:78}. Conversely, if the plane is tangent with some
$\left(n-k-1\right)$-dimensional boundary, it does not imply that
the tangent point must be the optimum point though, unless Eq. \eqref{eq:79}
is satisfied, which ensures the average success probability at any
other point in the feasible set will be no larger than that at the
tangent point, due to the convexity of the feasible set. This gives the geometrical meaning of Theorem \ref{boundary}.

\begin{figure}
\includegraphics[scale=0.5]{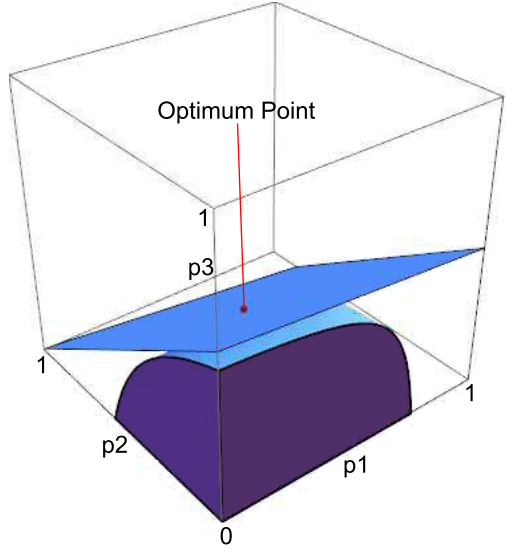}

\caption{\label{fig:tangent}(Color online)The optimum point is a non-singular interior point of the critical feasible region.}

\end{figure}

\begin{figure}
\includegraphics[scale=0.5]{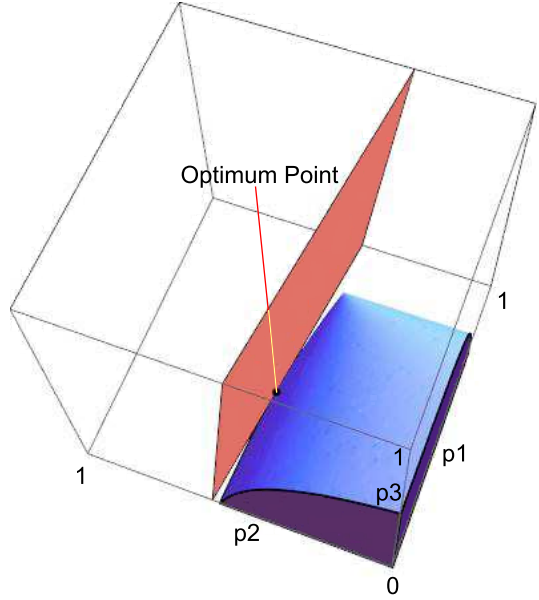}

\caption{\label{fig:boundary}(Color online)The optimum point is on the boundary $\mathcal{B}_{R}$.}

\end{figure}

\begin{figure}
\includegraphics[scale=0.5]{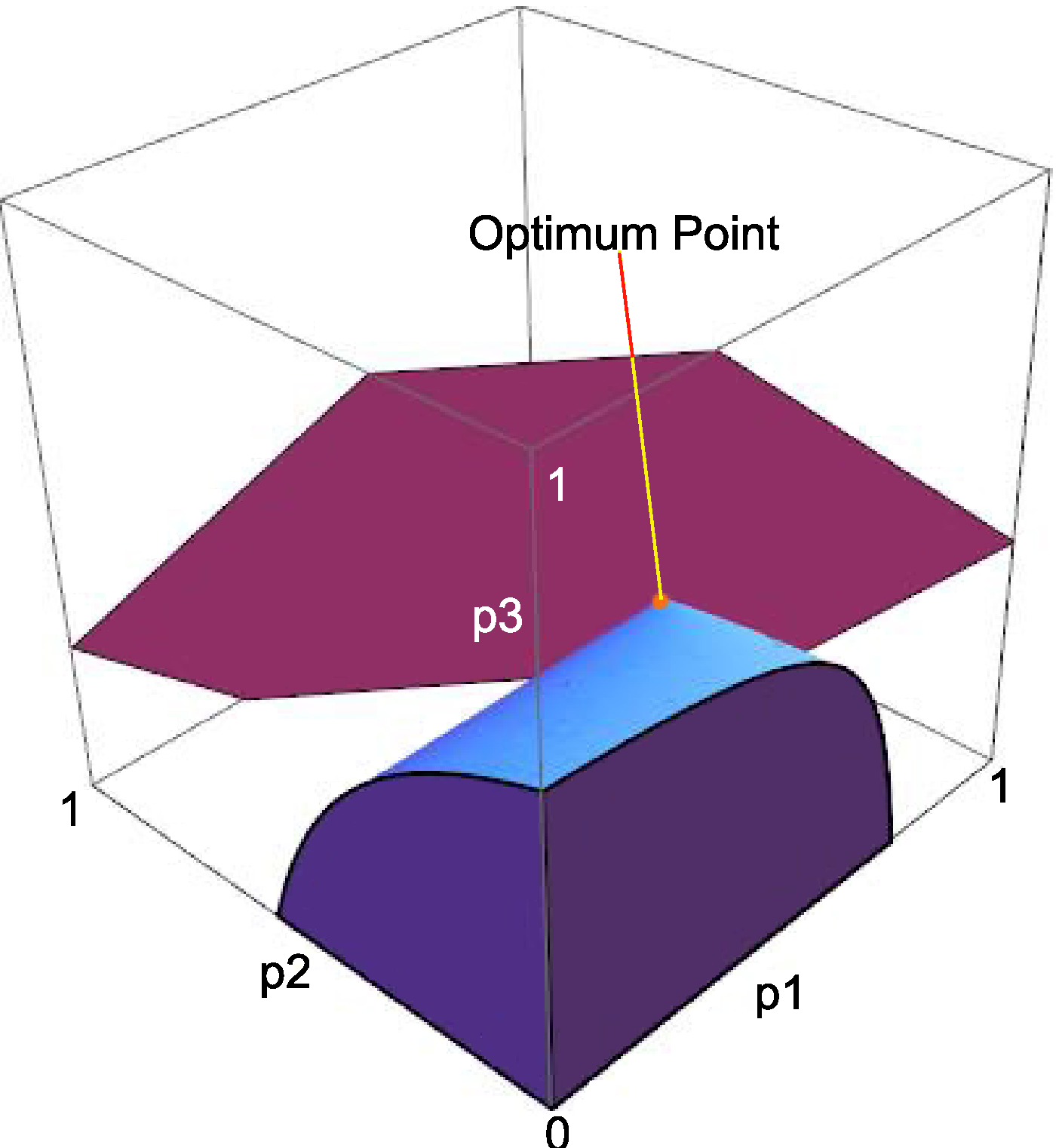}

\caption{\label{fig:singular}(Color online)The optimum point is a singular point.}

\end{figure}

From above, it can be seen that given the states to be discriminated, the category of the optimum point
is not determined: it may be a non-singular interior point in the critical feasible region,
a point on the boundary of the critical feasible region, or even a singular point, depending on
the prior probabilities. Taking the states in \eqref{eq:66} as an example,
we numerically obtain the optimum points for three different sets
of prior probabilities which  result in the above three different categories of the optimum points.
The results are presented in Table \ref{tab:numerical table} and  the corresponding graphics
are Fig. \ref{fig:tangent}, Fig. \ref{fig:boundary} and Fig. \ref{fig:singular}
respectively.
\begin{table}
\caption{\label{tab:numerical table}Numerical results of optimum points with different prior probabilities}
\begin{tabular}{ccccccc}
\hline
Category & Prior Probabilities & $p_{1}$ & $p_{2}$ & $p_{3}$ & $\lambda$ & $\bar{p}_{opt}$\tabularnewline
\hline
\hline
Interior Point & $\gamma_{1}=0.05,\gamma_{2}=0.35,\gamma_{3}=0.60$ & $0.5029$ & $0.3169$ & $0.3629$ & $0.2326$ & $0.3538$\tabularnewline
\hline
Boundary Point & $\gamma_{1}=0.10,\gamma_{2}=0.80,\gamma_{3}=0.10$ & $0.3927$ & $0.5300$ & $0$ & $0.6577$ & $0.4632$\tabularnewline
\hline
Singular Point & $\gamma_{1}=0.30,\gamma_{2}=0.35,\gamma_{3}=0.35$ & $0.6667$ & $0.4000$ & $0.2941$ & $0$ & $0.4429$\tabularnewline
\hline
\end{tabular}
\end{table}


\section{Analytical Relation between the  optimum solution and the states to be discriminated\label{sec:Maximum-Average-Success}}

In this section, we shall use the equations established in the previous section to make some efforts on the analytical optimum solution to the problem of unambiguously discriminating $n$ pure states. We shall obtain a formula which is not a complete analytical solution but can characterize a clear relation between the optimum solution and the states to be discriminated, and we shall give an example to show the use of that formula.

\subsection{Formula\label{sub:The-Analytical-Form}}

First of all, we give another form of the matrix $X-\Gamma$ in a way similar to Eqs. (2.5) and (2.6) of \cite{determinant 0}. Suppose
the states $|\psi_{k}\rangle$ ($k=1,\cdots,n$) are represented in
an orthonormal basis of the Hilbert space $\mathcal{H}$ spanned by
$\left\{ |\psi_{k}\rangle\right\} _{k=1}^{n}$, then each $|\psi_{k}\rangle$
has exactly $n$ components in its representation. If the initial
state of the system is $|\psi_{k}\rangle$, then the state after the inconclusive
measurement result can be chosen as\begin{equation}
|\phi_{k}\rangle=\sqrt{\Pi_{0}}|\psi_{k}\rangle,\; k=1\cdots n,\label{eq:90}\end{equation}
where $\Pi_{0}$ is defined in Eq. \eqref{eq:-9}. Define an $n\times n$
matrix $C$ with $|\phi_{k}\rangle$ as its $k$th column,\begin{equation}
C=\left(|\phi_{1}\rangle,\cdots,|\phi_{n}\rangle\right).\label{eq:48}\end{equation}
It can be directly verified that\begin{equation}
X-\Gamma=C^{\dagger}C,\label{eq:49}\end{equation}
using Eq. \eqref{eq:-25}.

Before presenting the main theorem of this section, we give three lemmas that will be used later as follows.

\refstepcounter{lem}\label{rank}\emph{Lemma \thelem.} Suppose $A$ is an $n\times n$ matrix. Let
the adjugate of $A$ be denoted by $A^{\ast}$, whose $\left(i,j\right)$
entry is the $\left(j,i\right)$ cofactor of $A$. Then \begin{equation}
\mathrm{Rank}\: A^{\ast}=\begin{cases}
n & \quad\mathrm{if\ Rank}\: A=n\\
1 & \mathrm{\quad if\ Rank}\: A=n-1\\
0 & \quad\mathrm{if\ Rank}\: A\leq n-2\end{cases}.\label{eq:41}\end{equation}

\refstepcounter{lem}\label{binetcauchy}\emph{Lemma \thelem (Binet-Cauchy).} Suppose $A$ and $B$ are matrices
of sizes $m\times n$ and $n\times m$ respectively, then\begin{equation}
\det\left(AB\right)=\begin{cases}
\qquad0 & \text{if \ensuremath{m>n}}\\
\det A\cdot\det B & \text{if \ensuremath{m=n}}\\
\underset{1\leq j_{1}<\cdots<j_{m}\leq n}{\sum}\det A\begin{pmatrix}1 & \cdots & m\\
j_{1} & \cdots & j_{m}\end{pmatrix}\det B\begin{pmatrix}j_{1} & \cdots & j_{m}\\
1 & \cdots & m\end{pmatrix} & \text{if \ensuremath{m<n}}\end{cases}.\label{eq:7}\end{equation}
where $A\begin{pmatrix}1 & \cdots & m\\
j_{1} & \cdots & j_{m}\end{pmatrix}$ denotes the $m\times m$ submatrix of $A$ whose $k$th column is the $j_{k}$th
column of $A$, and $B\begin{pmatrix}j_{1} & \cdots & j_{m}\\
1 & \cdots & m\end{pmatrix}$ denotes the $m\times m$ submatrix of $B$ whose $k$th row is the $j_{k}$th
row of $B$.

Proof of these two lemmas can be found in many algebra text books.

\refstepcounter{lem}\label{cstar}\emph{Lemma \thelem.} Let $C^{\ast}$ denote the adjugate of the
matrix $C$ defined in Eq. \eqref{eq:48} and $|c_{k}\rangle$ denote
the transpose of the $k$th row of $C^{\ast}$ ($k=1,\cdots,n$).
If the optimum point $\boldsymbol{p}_{opt}$ is a non-singular interior
point in the critical feasible region, then at $\boldsymbol{p}_{opt}$
each $|c_{k}\rangle$ can be written as\begin{equation}
|c_{k}\rangle|_{\boldsymbol{p}_{opt}}=\sqrt{\gamma_{k}}\xi e^{i\theta_{k}}|\rho\rangle,\label{eq:50}\end{equation}
where $|\rho\rangle$ is some normalized vector, $\xi$ is a
positive parameter and $e^{i\theta_{k}}$ is a phase to be determined.

\emph{Proof.} Let $C_{k}$ denote the submatrix of $C$ by deleting
the $k$th column $|\phi_{k}\rangle$, and $\left(X-\Gamma\right)_{ij}$
denote the submatrix of $X-\Gamma$ by deleting the $i$th row and
the $j$th column. Similar to Eq. \eqref{eq:49}, it can be directly verified that\begin{equation}
\left(X-\Gamma\right)_{ij}=C_{i}^{\dagger}C_{j},\label{eq:42}\end{equation}
so\begin{equation}
\det\left(X-\Gamma\right)_{ij}=\det\bigl(C_{i}^{\dagger}C_{j}\bigr)=\left(-1\right)^{i+j}\langle c_{i}|c_{j}\rangle\label{eq:43}\end{equation}
according to Lemma \ref{binetcauchy}.

Since $\boldsymbol{p}_{opt}$ is a non-singular interior point in
the critical feasible region $\mathcal{R}_{CF}$, we have $M_{k}\left(\boldsymbol{p}_{opt}\right)\neq0$
for some $k\in\left\{ 1,\cdots,n\right\} $, which implies that $\mathrm{Rank}\left(X-\Gamma\right)|_{\boldsymbol{p}_{opt}}=n-1$, or $\mathrm{Rank}\: C|_{\boldsymbol{p}_{opt}}=n-1$, equivalently.
Therefore, $\mathrm{Rank}\: C^{\ast}|_{\boldsymbol{p}_{opt}}=1$ according
to Lemma \ref{rank}, implying that all $|c_{k}\rangle$'s are proportional
to each other. So each $|c_{k}\rangle$ at $\boldsymbol{p}_{opt}$
can be written as\begin{equation}
|c_{k}\rangle|_{\boldsymbol{p}_{opt}}=a_{k}|\rho\rangle,\label{eq:54}\end{equation}
where $|\rho\rangle$ is some normalized vector and $a_{k}$ is a
coefficient to be determined.

From Eq. \eqref{eq:43}, we have\begin{equation}
M_{k}\left(\boldsymbol{p}\right)=\det\left(X-\Gamma\right)_{kk}=\langle c_{k}|c_{k}\rangle.\label{eq:33}\end{equation}
By substituting Eqs. \eqref{eq:5} and \eqref{eq:54} into \eqref{eq:33},
we have\begin{equation}
a_{k}^{\ast}a_{k}=\gamma_{k} \lambda,\label{eq:55}\end{equation} where $\lambda>0$.
Let $\lambda=\xi^{2}$, then Eq. \eqref{eq:50} holds.

The factor $\xi$ cannot be zero, otherwise $\sigma_{n}\left(\boldsymbol{p}\right)=0$
would be degenerate at $\boldsymbol{p}_{opt}$, contradicting the
assumption that $\boldsymbol{p}_{opt}$ is not a singular point. $\blacksquare$

When the states to be discriminated and the prior probabilities are
fixed, the difference between any pair of phase factors $e^{i\theta_{i}}$,
$e^{i\theta_{j}}$ $\left(i\neq j\right)$ is fixed, while the phase
factors $e^{i\theta_{k}}$ ($k=1,\cdots,n$) themselves can be altered since an arbitrary
total phase can always be added to all $e^{i\theta_{k}}$'s
by choosing an appropriate phase for $|\rho\rangle$. The phase differences
are what really matter in the following discussion.

\refstepcounter{thm}\label{analsol}\emph{Theorem \thethm.} If the optimum point $\boldsymbol{p}_{opt}$ is
a non-singular point in the interior part $\Omega_{R}$ of the critical
feasible region, then the components of $\boldsymbol{p}_{opt}$ can
be written as\begin{equation}
p_{i}=e^{-i\theta_{i}}\sum_{k=1}^{n}e^{i\theta_{k}}\sqrt{\frac{\gamma_{k}}{\gamma_{i}}}\langle\psi_{i}|\psi_{k}\rangle,
\quad\forall i=1,\cdots,n,\label{eq:44}\end{equation}
and the optimum average success probability can also be written as\begin{equation}
\bar{p}_{opt}=\biggl\Vert \sum_{k=1}^{n}\sqrt{\gamma_{k}}e^{i\theta_{k}}|\psi_{k}\rangle\biggr\Vert ^{2}.\label{eq:11}\end{equation}

\emph{Proof.} Since $\boldsymbol{p}_{opt}$ is a non-singular interior
point in the critical feasible region, by substituting Eq. \eqref{eq:50}
in Lemma \ref{cstar} into \eqref{eq:43}, we get\begin{equation}
\det\bigl(C_{i}^{\dagger}C_{k}\bigr)|_{\boldsymbol{p}_{opt}}=\left(-1\right)^{i+k}\sqrt{\gamma_{i}\gamma_{k}}e^{-i\theta_{i}+i\theta_{k}}\xi^{2},\;\forall i,k=1,\cdots,n.\label{eq:8}\end{equation}
Noting that the $(i,j)$ entry of $X-\Gamma$ is $\langle\psi_i|\psi_j\rangle-p_i\delta_{ij}$, the algebraic cofactor of the $(i,j)$ entry is $(-1)^{i+j}\det\left(X-\Gamma\right)_{ij}$ and $\det(X-\Gamma)=0$, we perform Laplace expansion on the determinant of $X-\Gamma$
along its $i$th row,\begin{equation}\begin{split}
\mathrm{\det}\left(X-\Gamma\right)|_{\boldsymbol{p}_{opt}} & =\sum_{k=1}^{n}(-1)^{i+k}\langle\psi_{i}|\psi_{k}\rangle\det\left(X-\Gamma\right){}_{ik}|_{\boldsymbol{p}_{opt}}-p_{i}M_{i}\left(\boldsymbol{p}_{opt}\right) \\
 & =\sum_{k=1}^{n}(-1)^{i+k}\langle\psi_{i}|\psi_{k}\rangle\det\bigl(C_{i}^{\dagger}C_{k}\bigr)|_{\boldsymbol{p}_{opt}}-p_{i}M_{i}\bigl(\boldsymbol{p}_{opt}\bigr) \\
 & =\sum_{k=1}^{n}\langle\psi_{i}|\psi_{k}\rangle\sqrt{\gamma_{i}\gamma_{k}}e^{-i\theta_{i}+i\theta_{k}}\xi^{2}-\gamma_{i}p_{i}\xi^{2}=0,\end{split}\label{eq:9}\end{equation}
for $i=1,\cdots,n$, where we have used Eqs. \eqref{eq:43}, \eqref{eq:33} and \eqref{eq:8}. Eliminating
$\xi^{2}$ from both sides of \eqref{eq:9}, we obtain\begin{equation}
\gamma_{i}p_{i}=\sum_{k=1}^{n}\langle\psi_{i}|\psi_{k}\rangle\sqrt{\gamma_{i}\gamma_{k}}e^{-i\theta_{i}+i\theta_{k}}=\sqrt{\gamma_{i}}e^{-i\theta_{i}}\langle\psi_{i}|\sum_{k=1}^{n}\sqrt{\gamma_{k}}e^{i\theta_{k}}|\psi_{k}\rangle,\quad\forall i,k=1,\cdots,n,\label{eq:10}\end{equation}
which immediately implies Eq. \eqref{eq:44}. Summing up Eq. \eqref{eq:10}
for all $i=1,\cdots,n$, we eventually get Eq. \eqref{eq:11}. $\blacksquare$

It should be pointed out that a similar result was derived in Ref.
\cite{08 nian} for the special case where $X$ is a real matrix, and our
Theorem \ref{analsol} can be considered as a generalization of that result to
the situation where $X$ is complex.

It should also be mentioned that Theorem \ref{analsol} actually gives an analytical relation
between the maximum average success probability and the $n$ pure
states to be discriminated but not a complete analytical solution,
since the explicit expressions of the phases $e^{i\theta_{k}}$ ($k=1,\cdots,n$) are not given in Theorem \ref{analsol}. However, Theorem
\ref{analsol} may sometimes help to simplify the calculation of the optimum solution
in special cases, as we shall show in the next subsection. And it
may also help to obtain some bounds of $\bar{p}_{opt}$ or work out
the phases $e^{i\theta_{k}}$ ($k=1,\cdots,n$) by numerical methods
according to Remark \ref{remark:extreme} given later.

\refstepcounter{rmk}\label{remark:phase}\emph{Remark \thermk.}  It seems that the solution \eqref{eq:44} and \eqref{eq:11}
would change if the phase of any state $|\psi_{k}\rangle$
is changed. But actually the corresponding $e^{i\theta_{k}}$
will also be changed in that case and we can see below that any term $e^{i\theta_{k}}|\psi_{k}\rangle$
in Eqs. \eqref{eq:44} and \eqref{eq:11} remains unchanged up to a global phase for all $e^{i\theta_{k}}|\psi_{k}\rangle$'s.

In fact, if some $|\psi_{i}\rangle$ is
transformed as\begin{equation}
|\psi_{i}\rangle\rightarrow e^{i\chi}|\psi_{i}\rangle,\label{eq:69}\end{equation}
where $e^{i\chi}$ is an arbitrary phase while the other $n-1$ states
stay unchanged, then according to the definitions of $|c_{k}\rangle$ and
$e^{i\theta_{k}}$ in Lemma \ref{cstar}, the phase $e^{i\theta_{k}}$ changes as follows:\begin{equation}
e^{i\theta_{k}}\rightarrow\begin{cases}
e^{i\left(\theta_{k}+\chi\right)} & \qquad\text{if }k\neq i\\
e^{i\theta_{k}}\text{ (unchanged)} & \qquad\text{if }k=i\end{cases}.\label{eq:70}\end{equation}
Since any global phase can be eliminated from all $e^{i\theta_{k}}$'s
by the vector $|\rho\rangle$, Eq. \eqref{eq:70} is equivalent to\begin{equation}
e^{i\theta_{k}}\rightarrow\begin{cases}
e^{i\theta_{k}}\text{ (unchanged)} & \qquad\textrm{if }k\neq i\\
e^{i\left(\theta_{k}-\chi\right)} & \qquad\textrm{if }k=i\end{cases}.\label{eq:71}\end{equation}
Therefore any term $e^{i\theta_{k}}|\psi_{k}\rangle$ (including the one that $k=i$) stays unchanged, considering Eq. \eqref{eq:69}.

\refstepcounter{rmk}\label{remark:extreme}\emph{Remark \thermk.} The maximum average probability  has an interesting property that it must be the value of a stationary point \cite{stationary point} of the
expression at the right side of Eq. \eqref{eq:11} if the phases
$e^{i\theta_{k}}$ are allowed to change freely. This is because
$p_{i}$ is  real  which requires that\begin{equation}
\sqrt{\gamma_{i}}e^{-i\theta_{i}}\langle\psi_{i}|\biggl(\sum_{k=1}^{n}\sqrt{\gamma_{k}}e^{i\theta_{k}}|\psi_{k}\rangle\biggr)-\biggl(\sum_{k=1}^{n}\sqrt{\gamma_{k}}e^{-i\theta_{k}}\langle\psi_{k}|\biggr)\sqrt{\gamma_{i}}e^{i\theta_{i}}|\psi_{i}\rangle=0,\label{eq:12}\end{equation}
according to Eq. \eqref{eq:44}, and  Eq. \eqref{eq:12} is equivalent to\begin{equation}
\frac{\partial}{\partial\theta_i}\biggl\Vert \sum_{k=1}^{n}\sqrt{\gamma_{k}}e^{i\theta_{k}}|\psi_{k}\rangle\biggr\Vert ^{2}=0, \label{eq:53}
\end{equation}
which is exactly the restriction equation of $\theta_{i}$
that must be satisfied when the expression at the right side of \eqref{eq:11} reaches a stationary point.

\subsection{Example\label{sub:An-Example}}

In this subsection, we give an example to show the use of Theorem
\ref{analsol}.

Suppose $\left\{ |\psi_{i}\rangle\right\} _{i=1}^{n}$ is a set
of linearly independent pure states that $\langle\psi_{1}|\psi_{i}\rangle\neq0$,
$i=2,\cdots,n$ and $\langle\psi_{i}|\psi_{j}\rangle=0$, $\forall i,j=2,\cdots,n$,
let's calculate the maximum average success probability when the optimum
point $\boldsymbol{p}_{opt}$ is a non-singular interior point in
the critical feasible region $\mathcal{R}_{CF}$.

Without loss of generality, we can choose that $e^{i\theta_{1}}=1$.
According to Eq. \eqref{eq:44} and the fact that
$p_{i}$ is real ($i=1,\cdots,n$), we can directly get\begin{equation}
\theta_{k}=\mathrm{Arg}\left(\langle\psi_{k}|\psi_{1}\rangle\right)-\pi,\quad (k=2,\cdots,n),\label{eq:63}\end{equation}
so\begin{equation}
\begin{cases}
p_{k}=1-\sqrt{\frac{\gamma_{1}}{\gamma_{k}}}|\langle\psi_{k}|\psi_{1}\rangle| \qquad (k=2,\cdots,n)\\
p_{1}=1-\sum_{k=2}^{n}\sqrt{\frac{\gamma_{k}}{\gamma_{1}}}|\langle\psi_{1}|\psi_{k}\rangle|\end{cases}.\label{eq:39}\end{equation}
Substituting Eq. \eqref{eq:63} into \eqref{eq:11} or using $\bar{p}=\sum_{i=1}^{n}\gamma_{i}p_{i}$,
we have\begin{equation}
\bar{p}_{opt}=\sum_{k=1}^{n}\gamma_{k}p_{k}=1-2\sum_{k=2}^{n}\sqrt{\gamma_{1}\gamma_{k}}|\langle\psi_{k}|\psi_{1}\rangle|.\label{eq:45}\end{equation}

Since $0\leq p_{k}\leq1$, $k=1,\cdots,n$, according to Eq. \eqref{eq:39},
we obtain $\sqrt{\frac{\gamma_{k}}{\gamma_{1}}}\geq|\langle\psi_{k}|\psi_{1}\rangle|$
(for all $k=2,\cdots,n$) and $\sum_{k=2}^{n}\sqrt{\frac{\gamma_{k}}{\gamma_{1}}}|\langle\psi_{1}|\psi_{k}\rangle|\leq1$,
which are the conditions for the optimum point $\boldsymbol{p}_{opt}$ to be a non-singular
interior point in the critical feasible region $\mathcal{R}_{CF}$.

When $n=2$, Eq. \eqref{eq:45} gives the well known Ivanovic-Dieks-Peres
limit \cite{Dieks,Ivanovic,jaeger unambiguous,Peres}.

\section{A generalized equal-probability measurement problem\label{sec:Generalization-of-Equal-Probability}}

A special scheme to discriminate quantum states unambiguously is the
so-called \emph{equal-probability measurement} (EPM) \cite{ieee epm},
which requires that the probability of each measurement outcome is
equal, i.e., $p_{1}=\cdots=p_{n}$. In this section we will treat
a generalized version of the EPM problem, which is defined as follows.

\emph{The Generalized EPM Problem (GEPM)}. If it is required that\begin{equation}
p_{1}:p_{2}:\cdots:p_{n}=w_{1}:w_{2}:\cdots:w_{n}\label{eq:47}\end{equation}
for a given set of non-negative numbers $w_{i}$ ($w_{i}\geq0,\forall i=1,\cdots,n$)
when the average success probability of unambiguously discriminating
the states $\left\{ |\psi_{i}\rangle\right\} _{i=1}^{n}$ reaches
the maximum, one needs to work out the prior probabilities $\left\{ \gamma_{i}\right\} _{i=1}^{n}$
or the conditions these prior probabilities should satisfy.

We have the following result on this generalized version of EPM problem
using Theorem \ref{nonsigular}.

\refstepcounter{thm}\label{GEPM}\emph{Theorem \thethm (Generalized EPM).} Suppose $\Psi$ is a matrix with
$|\psi_{i}\rangle/\sqrt{w_{i}}$ as its $i$th column, let $\sigma_{min}$
denote the minimum eigenvalue of $\Psi^{\dagger}\Psi$ and $M_{i}^{GEPM}=M_{i}\left(\boldsymbol{p}\right)|_{p_{1}=w_{1}\sigma_{min},\cdots,p_{n}=w_{n}\sigma_{min}}$
be $\left(n-1\right)\times\left(n-1\right)$ principal minor of $X-\Gamma$
corresponding to its $i$th diagonal element while $p_{1}=w_{1}\sigma_{min},\cdots,p_{n}=w_{n}\sigma_{min}$.
If $M_{i}^{GEPM}>0$ for some $i\in\left\{ 1,\cdots,n\right\} $,
then the sufficient and necessary conditions that the GEPM is the
optimum POVM to unambiguously discriminate the given states is that\begin{equation}
\gamma_{i}=\frac{M_{i}^{GEPM}}{\sum_{i=1}^{n}M_{i}^{GEPM}},\quad\forall i=1,\cdots,n.\label{eq:46}\end{equation}
If $M_{i}^{GEPM}=0$ for all $i\in\left\{ 1,\cdots,n\right\} $,
then there exist a range of different $\boldsymbol{\gamma}$'s for
which the GEPM is the optimum POVM for unambiguous discrimination
of the states $\left\{ |\psi_{i}\rangle\right\} _{i=1}^{n}$.

\emph{Proof.} We suppose that the optimum solution $\boldsymbol{p}_{opt}$
is $p_{1}=w_{1}\eta,\cdots,p_{n}=w_{n}\eta$ according to Eq.
\eqref{eq:47}, and the corresponding $\Gamma$ matrix is $\text{diag}(
w_{1}\eta,w_{2}\eta,\cdots,w_{n}\eta)$, where $\eta$ is to be determined. By some simple calculation,
it can be shown that in this situation the condition $X-\Gamma\geq0$ can be converted
to $\Psi^{\dagger}\Psi-\eta I\geq0$ where $I$ is the identity
matrix, so $\sigma_{min}$, the minimum eigenvalue
of $\Psi^{\dagger}\Psi$, is exactly the maximum feasible value of
$\eta$, which means that $p_{i}=w_{i}\sigma_{min}$ ($i=1,\cdots,n$)
is the optimum solution $\boldsymbol{p}_{opt}$ when Eq. \eqref{eq:47}
has to be satisfied.
If $M_{i}^{GEPM}\neq0$ for some $i\in\left\{ 1,\cdots,n\right\} $,
$\boldsymbol{p}_{opt}$ is not a singular point, so Eq. \eqref{eq:46}
holds according to Eq. \eqref{eq:5}.

On the other hand, if $M_{i}^{GEPM}=0$ for all $i\in\left\{ 1,\cdots,n\right\} $,
then the point where $p_{1}=w_{1}\sigma_{min},\cdots,p_{n}=w_{n}\sigma_{min}$
in the critical feasible region is a singular point, so the normal
vector changes discontinuously in the neighborhood of $\boldsymbol{p}=\left(w_{1}\sigma_{min},\cdots,w_{n}\sigma_{min}\right)$
in the critical feasible region $\mathcal{R}_{CF}$. Thus, the GEPM
is the optimum unambiguous discrimination scheme for a range of different
$\boldsymbol{\gamma}$'s. $\blacksquare$

The solution of the original EPM problem follows immediately from
Theorem \ref{GEPM} by setting $w_{1}=\cdots=w_{n}=1$.

It is obvious that for any set of linearly independent quantum states,
it is always possible to find prior probabilities $\left\{ \gamma_{1},\cdots,\gamma_{n}\right\} $
such that the generalized EPM is the optimum scheme to unambiguously
discriminate these states.

\section{Unambiguous discrimination of three pure states\label{sec:Special-Case:}}

In this section we shall use the results and method presented in Sec.
III to study the unambiguous discrimination problem of three linearly
independent pure states, mainly for the non-singular interior optimum solution.

\subsection{General equations\label{sub:General-Equations}}

Suppose the three states to be discriminated are $|\psi_{1}\rangle,|\psi_{2}\rangle,|\psi_{3}\rangle$
with prior probabilities $\gamma_{1},\gamma_{2},\gamma_{3}$, and
they are linearly independent. Then according to Eq. \eqref{eq:5}
we can have the following equations\begin{equation}
M_{1}\left(\boldsymbol{p}\right)=(1-p_{2})(1-p_{3})-|\langle\psi_{2}|\psi_{3}\rangle|^{2}=\lambda\gamma_{1},\label{eq:13}\end{equation}
\begin{equation}
M_{2}\left(\boldsymbol{p}\right)=(1-p_{1})(1-p_{3})-|\langle\psi_{1}|\psi_{3}\rangle|^{2}=\lambda\gamma_{2},\label{eq:14}\end{equation}
\begin{equation}
M_{3}\left(\boldsymbol{p}\right)=(1-p_{1})(1-p_{2})-|\langle\psi_{1}|\psi_{2}\rangle|^{2}=\lambda\gamma_{3}.\label{eq:15}\end{equation}

From these three equations, we can obtain \begin{equation}
1-p_{1}=\sqrt{\frac{\left(|\langle\psi_{1}|\psi_{2}\rangle|^{2}+\lambda\gamma_{3}\right)\left(|\langle\psi_{1}|\psi_{3}\rangle|^{2}+\lambda\gamma_{2}\right)}{|\langle\psi_{2}|\psi_{3}\rangle|^{2}+\lambda\gamma_{1}}},\label{eq:16}\end{equation}
\begin{equation}
1-p_{2}=\sqrt{\frac{\left(|\langle\psi_{2}|\psi_{3}\rangle|^{2}+\lambda\gamma_{1}\right)\left(|\langle\psi_{1}|\psi_{2}\rangle|^{2}+\lambda\gamma_{3}\right)}{|\langle\psi_{1}|\psi_{3}\rangle|^{2}+\lambda\gamma_{2}}},\label{eq:17}\end{equation}
\begin{equation}
1-p_{3}=\sqrt{\frac{\left(|\langle\psi_{2}|\psi_{3}\rangle|^{2}+\lambda\gamma_{1}\right)\left(|\langle\psi_{1}|\psi_{3}\rangle|^{2}+\lambda\gamma_{2}\right)}{|\langle\psi_{1}|\psi_{2}\rangle|^{2}+\lambda\gamma_{3}}}.\label{eq:18}\end{equation}

Substituting Eqs. \eqref{eq:16}, \eqref{eq:17} and \eqref{eq:18}
into $\det\left(X-\Gamma\right)=0$ and making some rearrangements,
we can get the following equation of $\lambda$\begin{equation}
\gamma\lambda^{3}-S\lambda-2|T|^{2}+2\sqrt{\gamma\lambda^{3}+R\lambda^{2}+S\lambda+|T|^{2}}\mathrm{Re}\left(T\right)=0,\label{eq:19}\end{equation}
where $\mathrm{Re}\left(T\right)$ represents the real part of $T$ and $S,\:T,\:R,\:\gamma$ are defined as
\begin{equation}
\gamma=\gamma_{1}\gamma_{2}\gamma_{3}, \qquad\qquad T=\langle\psi_{1}|\psi_{2}\rangle\langle\psi_{2}|\psi_{3}\rangle\langle\psi_{3}|\psi_{1}\rangle,\label{eq:20}\end{equation}
\begin{equation}
R=\gamma_{1}\gamma_{2}|\langle\psi_{1}|\psi_{2}\rangle|^{2}+\gamma_{2}\gamma_{3}|\langle\psi_{2}|\psi_{3}\rangle|^{2}+\gamma_{1}\gamma_{3}|\langle\psi_{1}|\psi_{3}\rangle|^{2},\label{eq:21}\end{equation}
\begin{equation}
S=\gamma_{1}|\langle\psi_{1}|\psi_{2}\rangle|^{2}|\langle\psi_{3}|\psi_{1}\rangle|^{2}+\gamma_{2}|\langle\psi_{1}|\psi_{2}\rangle|^{2}|\langle\psi_{2}|\psi_{3}\rangle|^{2}+\gamma_{3}|\langle\psi_{2}|\psi_{3}\rangle|^{2}|\langle\psi_{3}|\psi_{1}\rangle|^{2}.\label{eq:22}\end{equation}

In Ref. \cite{chefles unambiguous 1}, it is doubted that whether
a general closed form of the maximum average success probability
of unambiguous state discrimination exists.
One can show that Eq. \eqref{eq:19} can be converted to a polynomial equation of degree $6$,
so it is generally difficult to find an analytical
solution for unambiguous discrimination of three states. However,
in some special situations, analytical solutions can be obtained,
and we will give some examples in the next two subsections.

\subsection{Special cases\label{sub:Special-Cases}}

For some special cases, Eq. \eqref{eq:19} can be simplified and one
can obtain exact analytical solutions.

\emph{Case 1}. Suppose that $\langle\psi_{1}|\psi_{2}\rangle=0$, but
$\langle\psi_{2}|\psi_{3}\rangle\neq0$, $\langle\psi_{3}|\psi_{1}\rangle\neq0$,
then Eq. \eqref{eq:19} can be simplified to\begin{equation}
\gamma_{1}\gamma_{2}\gamma_{3}\lambda^{3}-|\langle\psi_{3}|\psi_{1}\rangle\langle\psi_{2}|\psi_{3}\rangle|^{2}\gamma_{3}\lambda=0.\label{eq:23}\end{equation}
We can easily obtain that $\lambda=0$ or\begin{equation}
\lambda=\frac{|\langle\psi_{3}|\psi_{1}\rangle\langle\psi_{2}|\psi_{3}\rangle|}{\sqrt{\gamma_{1}\gamma_{2}}},\label{eq:68}\end{equation}
and the negative root has been discarded.

With some observation, it can be verified that $\lambda=0$ is not
a solution for Eqs. \eqref{eq:16}-\eqref{eq:18}, unless $\langle\psi_{1}|\psi_{3}\rangle=0$
or $\langle\psi_{2}|\psi_{3}\rangle=0$, which is a trivial case.
Thus Eq. \eqref{eq:68} is the unique solution for this situation.
Substituting Eq. \eqref{eq:68} into \eqref{eq:16}-\eqref{eq:18},
we can get the optimum average probability\begin{align}
\overline{p}_{opt} & =1-2\sqrt{\gamma_{1}\gamma_{3}}|\langle\psi_{1}|\psi_{3}\rangle|-2\sqrt{\gamma_{2}\gamma_{3}}|\langle\psi_{2}|\psi_{3}\rangle|.\label{eq:24}\end{align}
This agrees with the result in Sec. \ref{sub:An-Example}.

\emph{Case 2}. Suppose that $\langle\psi_{1}|\psi_{2}\rangle\langle\psi_{2}|\psi_{3}\rangle\langle\psi_{3}|\psi_{1}\rangle$
is purely imaginary, i.e. $\mathrm{Re}\left(\mbox{\ensuremath{\langle\psi_{1}}|\ensuremath{\psi_{2}\rangle\langle\psi_{2}}|\ensuremath{\psi_{3}\rangle\langle\psi_{3}}|\ensuremath{\psi_{1}\rangle}}\right)=0$,
but $\mbox{\ensuremath{\langle\psi_{1}}|\ensuremath{\psi_{2}\rangle\langle\psi_{2}}|\ensuremath{\psi_{3}\rangle\langle\psi_{3}}|\ensuremath{\psi_{1}\rangle}}\neq0$,
then Eq. \eqref{eq:19} becomes\begin{equation}
\gamma\lambda^{3}-S\lambda-2|T|^{2}=0.\label{eq:25}\end{equation}
This is a cubic equation of $\lambda$, so we can get analytical solutions
in general. It can be proved that this equation has and only
has one positive root of $\lambda$ using Vieta's theorem \cite{vieta's theorem},
and this positive root is\begin{align}
\lambda & =\frac{\left(27\gamma^{2}|T|^{2}+3\sqrt{81\gamma^{4}|T|^{4}-3\gamma^{3}S^{3}}\right)^{\frac{1}{3}}}{3\gamma}+\frac{S}{\left(27\gamma^{2}|T|^{2}+3\sqrt{81\gamma^{4}|T|^{4}-3\gamma^{3}S^{3}}\right)^{\frac{1}{3}}},\label{eq:26}\end{align}
where we have used the notations defined in Eqs. \eqref{eq:20}-\eqref{eq:22}.

It can be shown that\begin{equation}
S^{3}\geq27\gamma|T|^{4}\label{eq:56}\end{equation}
by the mean inequality \cite{mean inequality}, so Eq. \eqref{eq:26}
can be simplified to\begin{equation}
\lambda=2\sqrt{\frac{S}{3\gamma}}\cos\frac{\theta}{3},\label{eq:27}\end{equation}
where\begin{equation}
\theta=\arccos\frac{|T|^{2}}{S}\sqrt{\frac{27\gamma}{S}}.\label{eq:28}\end{equation}

Substituting Eq. \eqref{eq:27} into \eqref{eq:16}-\eqref{eq:18},
we can eventually get $p_{1}$, $p_{2}$, $p_{3}$ and $\bar{p}$ for
the optimum solution.

\emph{Case 3.} If $\langle\psi_{1}|\psi_{2}\rangle\langle\psi_{2}|\psi_{3}\rangle\langle\psi_{3}|\psi_{1}\rangle$
is real and non-negative, then it can be directly verified that $\lambda=0$
is a solution to Eq. \eqref{eq:19}, which implies that a singular point exists in the critical feasible region. And in this case, Eq. \eqref{eq:19}
can be simplified to\begin{equation}
\lambda^{2}\left(\gamma^{2}\lambda^{4}-2\gamma S\lambda^{2}-8\gamma|T|^{2}\lambda-Q\right)=0,\label{eq:30}\end{equation}
where in addition to Eqs. \eqref{eq:20}-\eqref{eq:22}, another constant
$Q$ is defined as\begin{equation}
Q=S^{2}-4R|T|^{2}.\label{eq:31}\end{equation}

Since the expression inside the parentheses of Eq. \eqref{eq:30}
is a quadratic polynomial, analytical solutions of $\lambda$ can
be obtained from \eqref{eq:30}. But the analytical solution
is too complicated to show any practical meaning, so we are not going
to include it here.

It should be pointed out that the analytical solutions for three pure states with  with real Gram matrix
was given in Ref. \cite{08 nian}, and our Case 3 generalizes that
result.

\subsection{Equal-Probability Measurement\label{sub:Equal-Probability-Measurement}}

In Sec. \ref{sec:Generalization-of-Equal-Probability}, we have given
the exact solutions of the prior probabilities
to the generalized EPM problem. As an example, we solve the original
EPM problem  of three pure states in a direct way.

Let $p_{1}=p_{2}=p_{3}=p_{EPM}$ and substitute it into $\det\left(X-\Gamma\right)=0$,
one can get\begin{eqnarray}
p_{EPM} & = & 1-\frac{\left(1+3i\right)W}{2\cdot3^{\frac{1}{3}}\biggl(\sqrt{3}\sqrt{27\left(\mathrm{Re}T\right)^{2}-W^{3}}-9\mathrm{Re}T\biggr)^{\frac{1}{3}}}-\frac{\left(1-3i\right)\biggl(\sqrt{3}\sqrt{27\left(\mathrm{Re}T\right)^{2}-W^{3}}-9\mathrm{Re}T\biggr)^{\frac{1}{3}}}{2\cdot3^{2/3}},\label{eq:34}\end{eqnarray}
where $T$ is defined in Eq. \eqref{eq:20} and\begin{equation}
W=|\langle\psi_{1}|\psi_{2}\rangle|^{2}+|\langle\psi_{1}|\psi_{3}\rangle|^{2}+|\langle\psi_{2}|\psi_{3}\rangle|^{2}.\label{eq:38}\end{equation}

It can be verified that\begin{equation}
27\left(\mathrm{Re}T\right)^{2}\leq27|T|^{2}\leq W^{3},\label{eq:57}\end{equation}
where the second inequality can be proved by the mean inequality \cite{mean inequality},
then $p_{EPM}$ can be reduced to\begin{equation}
p_{EPM}=1-2\cdot\sqrt{\frac{W}{3}}\cos\biggl(\frac{\pi}{3}-\frac{\theta}{3}\biggr),\label{eq:35}\end{equation}
where\begin{equation}
\theta=\arccos\frac{3\sqrt{3}\mathrm{Re}\left(T\right)}{W\sqrt{W}}.\label{eq:36}\end{equation}

Thus according to Eqs. \eqref{eq:13}-\eqref{eq:15}, the prior probabilities
$\gamma_{1},\gamma_{2},\gamma_{3}$ must be\begin{equation}
\begin{cases}
\gamma_{1}=\frac{\frac{4}{3}W\cos^{2}\left(\frac{\pi}{3}-\frac{\theta}{3}\right)-|\langle\psi_{2}|\psi_{3}\rangle|^{2}}{4W\cos^{2}\left(\frac{\pi}{3}-\frac{\theta}{3}\right)-W}\\
\gamma_{2}=\frac{\frac{4}{3}W\cos^{2}\left(\frac{\pi}{3}-\frac{\theta}{3}\right)-|\langle\psi_{1}|\psi_{3}\rangle|^{2}}{4W\cos^{2}\left(\frac{\pi}{3}-\frac{\theta}{3}\right)-W}\\
\gamma_{3}=\frac{\frac{4}{3}W\cos^{2}\left(\frac{\pi}{3}-\frac{\theta}{3}\right)-|\langle\psi_{1}|\psi_{2}\rangle|^{2}}{4W\cos^{2}\left(\frac{\pi}{3}-\frac{\theta}{3}\right)-W}\end{cases}.\label{eq:37}\end{equation}

\refstepcounter{rmk}\label{three state epm}
\emph{Remark \thermk.}
In Eq. \eqref{eq:37}, the denominators of $\gamma_1$, $\gamma_2$, $\gamma_3$ can be equal to zero, and in this situation the optimum EPM point is a singular point. This can be shown as follows. If the dominators in \eqref{eq:37} are equal to zero, then\begin{equation} \cos\biggl(\frac{\pi}{3}-\frac{\theta}{3}\biggr)=\frac{1}{2},\label{eq:64}\end{equation}
so\begin{equation}
\theta=0 \;\;\text{or}\;\;2\pi.\label{eq:65}\end{equation}
(Note that $\cos\left(\frac{\pi}{3}-\frac{\theta}{3}\right)$ can not be $-\frac{1}{2}$, otherwise $p_{EPM}$ would be larger than $1$ according to Eq. \eqref{eq:35}.)\\
Then it can be seen that\begin{equation}
3\sqrt{3}\mathrm{Re}\left(T\right)=W\sqrt{W}\label{eq:29}\end{equation}
by substituting Eq. \eqref{eq:65} into \eqref{eq:36}.\\
According to Eq. \eqref{eq:57} and the condition that "=" holds in a mean inequality \cite{mean inequality}, one can have\begin{equation}
|\langle\psi_{1}|\psi_{2}\rangle|=|\langle\psi_{1}|\psi_{3}\rangle|=|\langle\psi_{2}|\psi_{3}\rangle|.\label{eq:4}\end{equation}
Then according to Eqs. \eqref{eq:38} and  \eqref{eq:35},\begin{equation}
p_1=p_2=p_3=p_{EPM}=1- |\langle\psi_{1}|\psi_{2}\rangle|\label{eq:58}.\end{equation}
Substituting Eq. \eqref{eq:58} into Eqs. \eqref{eq:13}-\eqref{eq:15}, it can be seen that\begin{equation}
M_{1}\left(\boldsymbol{p}\right)=M_{2}\left(\boldsymbol{p}\right)=M_{3}\left(\boldsymbol{p}\right)=0.\label{eq:6}\end{equation}
So, when the dominators in Eq. \eqref{eq:37} are equal to zero, the optimum EPM point is a singular point.

\section{Conclusion \label{sec:Conclusion-and-Perspective}}

In this paper, we have mainly studied the problem of optimum unambiguous
discrimination of $n$ linearly independent pure states. We have derived some analytical properties
of the optimum solution to this problem, and established
two sets of new equations in Theorems 3 and 4 which provide detailed methods
to obtain the optimum solution in different situations. We have also presented
a geometrical illustration of the equations we established with a numerical example in Sec. \ref{sub:Numerical-Example}.
An analytical formula which shows the relation between the optimum solution of the unambiguous discrimination problem and the $n$ pure states to be identified has been derived in Sec. \ref{sec:Maximum-Average-Success}.
And we have also solved a generalized EPM problem in Sec. \ref{sec:Generalization-of-Equal-Probability},
with the proportion of the occurring probabilities of the measurement outcomes to be fixed.
Finally, the optimum unambiguous discrimination
problem of three pure states is studied, and analytical results has been presented for
some interesting cases in Sec. \ref{sec:Special-Case:}.

It is no doubt that the problem of discriminating quantum states
is important in quantum information science since it has wide application
to quantum cryptography and quantum communication, so it motivates
a lot of researchers to explore different kinds of optimum discrimination strategies.
In addition to many important results mentioned in Sec. \ref{sec:Introduction},
some other novel strategies such as minimax discrimination \cite{minimax 1,minimax 2}
and maximum confidence discrimination \cite{max confidence 1,max confidence 2}
have been introduced recently. We hope that our results presented
in this article may stimulate further research to the optimum state
discrimination problem in general.

\begin{acknowledgments}
This research receives support from the NNSF of China (Grant No. 10604051),
the CAS, and the National Fundamental Research Program.
\end{acknowledgments}


\begin{thebibliography}{62}
\bibitem{cauchy schwartz}The Cauchy-Schwartz inequality: Let $\boldsymbol{x}$
and $\boldsymbol{y}$ be two arbitrary vectors in a Hilbert space,
then $\|\boldsymbol{x}\cdot\boldsymbol{y}\|\leq\sqrt{\|\boldsymbol{x}\|^{\frac{1}{2}}\|\boldsymbol{y}\|^{\frac{1}{2}}}$,
where $\|\cdot\|$ denotes the norm of a vector in the given
Hilbert space, and the "$=$" holds if and only if $\boldsymbol{x}$
and $\boldsymbol{y}$ are linearly dependent.
\bibitem{stationary point}A stationary point of a function $F(x_1,\cdots,x_n)$ is a point in the domain of the function where  $\frac{\partial F}{\partial x_i}=0,\, i=1,\cdots,n$.
\bibitem{vieta's theorem}Vieta's theorem characterizes the relation
between the coefficients and the roots of a polynomial equation as
follows: suppose $x^{n}+a_{1}x^{n-1}\cdots+a_{n-1}x+a_{n}=0$ is a
polynomial equation defined on the complex field $\mathbb{C}$ and
it has $n$ roots denoted by $x_{1},\cdots,x_{n}$, then there must
be $\sum_{1\leq i_{1}\leq\cdots\leq i_{k}\leq n}x_{i_{1}}x_{i_{2}}\cdots x_{i_{k}}=\left(-1\right)^{k}a_{k}$.
\bibitem{mean inequality}The mean inequality is $\frac{1}{n}\sum_{i=1}^{n}x_{i}\geq\left(\prod_{i=1}^{n}x_{i}\right)^{\frac{1}{n}}$ provided that $x_{1},\cdots,x_{n}$ are $n$ positive real numbers. And the condtion that "=" holds in the mean inequality is $x_1=\cdots=x_n$.
\end{thebibliography}
\end{document}